# Asynchronous Code-Division Random Access Using Convex Optimization

Lorne Applebaum, Waheed U. Bajwa, Marco F. Duarte, and Robert Calderbank

## Abstract

Many applications in cellular systems and sensor networks involve a random subset of a large number of users asynchronously reporting activity to a base station. This paper examines the problem of multiuser detection (MUD) in random access channels for such applications. Traditional orthogonal signaling ignores the random nature of user activity in this problem and limits the total number of users to be on the order of the number of signal space dimensions. Contention-based schemes, on the other hand, suffer from delays caused by colliding transmissions and the hidden node problem. In contrast, this paper presents a novel pairing of an asynchronous non-orthogonal code-division random access scheme with a convex optimization-based MUD algorithm that overcomes the issues associated with orthogonal signaling and contention-based methods. Two key distinguishing features of the proposed MUD algorithm are that it does not require knowledge of the delay or channel state information of every user and it has polynomial-time computational complexity. The main analytical contribution of this paper is the relationship between the performance of the proposed MUD algorithm in the presence of arbitrary or random delays and two simple metrics of the set of user codewords. The study of these metrics is then focused on two specific sets of codewords, random binary codewords and specially constructed algebraic codewords, for asynchronous random access. The ensuing analysis confirms that the proposed scheme together with either of these two codeword sets significantly outperforms the orthogonal signaling-based random access in terms of the total number of users in the system.

# I. INTRODUCTION

Many applications of wireless networks require servicing a large number of users that share limited communication resources. In particular, the term *random access* is commonly used to describe a setup where a random subset of users in the network communicate with a base station (BS) in an uncoordinated fashion [1]. In this paper, we study random access in large networks for the case when active users transmit single bits to the BS. This so-called "on–off" random access channel (RAC) [2] represents an abstraction that arises frequently in many wireless networks. In third-generation cellular systems, for example, control channels used for scheduling requests can be modeled as on–off RACs; in this case, users requesting permissions to send data to the BS can be thought of as transmitting 1's and inactive users can be thought of as transmitting 0's. Similarly, uplinks in wireless sensor networks deployed for target detection can also be modeled as on–off RACs; in this case, sensors that detect a target can be made to transmit 1's and sensors that have nothing to report can be thought of as transmitting 0's.[1]

The primary objective of the BS in on–off RACs is to reliably and efficiently carry out multiuser detection (MUD), which translates into recovery of the set of active users in our case. The two biggest impediments to this goal are that (*i*) random access tends to be asynchronous in nature, and (*ii*) it is quite difficult, if not impossible, for the BS to know the channel state information (CSI) of every user. Given a fixed number of temporal signal space dimensions $N$ in the uplink, the system-design goal therefore is to simultaneously maximize the total number of users $M$ in the network and the average number of active users $k$ that the BS can reliably handle *without* requiring knowledge of the delays or CSIs of the individual users at the BS.

Traditional approaches to random access fall significantly short of this design objective. In random access methods based on orthogonal signaling, the $N$ signal space dimensions are orthogonally spread among the $M$ users in either time, frequency, or code [1]. While this establishes a dedicated, interference-free channel between each user and the BS, this approach ignores the random nature of user activity in RACs. Therefore, by its very structure, random access based on orthogonal signaling dictates the relationship $k \leq M \leq N$. On the other hand, contention-based random access schemes such as ALOHA and carrier sense multiple access (CSMA) do take advantage of the random user activity [3]. However, significant problems arise in these schemes when the average number of active users $k$ and/or the total number of users $M$ gets large [3]. In the case of ALOHA, collisions and retransmissions accumulate to significant delays as $k$ becomes large. In the case of CSMA, the number of potential "hidden nodes" grows as $M$ increases, resulting in unintended and unrecognized collisions in large networks.

---

[1]The focus of this paper is on servicing a large number of users that share limited communication resources in the uplink. Limiting ourselves to on–off RACs in this case helps up isolate the key issues associated with designing arbitrary RACs involving (multiple-bit) packet transmissions in large networks.

Cellular systems, partly because of the aforementioned reasons, typically resort to the use of matched filter receivers on uplink control channels. Such receivers correspond to single-user detection (SUD) since they detect each user independently, treating the interference from other active users as noise. However, despite the effectiveness of these receivers in today's cellular systems, SUD schemes also have significant pitfalls. In particular, such schemes tend to have suboptimal performance since they do not carry out joint detection and they tend to be prone to the "near–far" effect [2].

In order to overcome the issues associated with orthogonal signaling, contention-based methods and SUD schemes, we present in this paper a novel code-division random access (CDRA) scheme that spreads the uplink communication resources in a non-orthogonal manner among the $M$ users and leverages the random user activity to service significantly more total users than $N$. A key distinguishing feature of the proposed scheme is that it makes use of a convex optimization-based MUD algorithm that does not require knowledge of the delays or CSIs of the users at the BS. In addition, we present an efficient implementation of the proposed algorithm based on the *fast Fourier transform* (FFT) that ensures that its computational complexity at worst differs by a logarithmic factor from an *oracle-based* MUD algorithm that has perfect knowledge of the user delays. Our main analytical contribution is the relationship between the probability of error $P_{err}$ of the proposed MUD algorithm in the presence of arbitrary or random delays and two metrics of the set of codewords assigned to the users. We then make use of these metrics to analyze two specific sets of codewords, random binary codewords and specially constructed (deterministic) algebraic codewords, for the proposed random access scheme. Specifically, we show that both these codewords allow our scheme to successfully manage an average number of active users that is almost linear in $N$: $k \lesssim N/(\tau \log(M\tau))$ for arbitrary delays and $k \lesssim N/\log(M\tau)$ for uniformly random delays, where $\tau$ denotes maximum delay in the network. More importantly, we show that the set of random codewords enable our scheme to service a number of total users that (ignoring $\tau$) is super-polynomial in $N$, $M \lesssim \exp(O(N^{1/3}))$, while the set of deterministic codes, which facilitate efficient codeword construction and storage, enable it to service a number of total users that is polynomial in $N$, $M \lesssim N^t$ for any reasonably sized $t \geq 2$.[2]

It is useful at this point to also consider non-orthogonal code-division *multiple access* (CDMA),

---

[2]Recall the "Big–O" notation: $f(n) = O(g(n))$ (alternatively, $f(n) \lesssim g(n)$) if $\exists\, c_o > 0, n_o : \forall\, n \geq n_o, f(n) \leq c_o g(n)$.

which—like our scheme—also spreads the uplink communication resources in a non-orthogonal manner among $M > N$ users [1]. However, despite similarities at the codeword-assignment level, there are significant differences between non-orthogonal CDMA and the work presented here. First, non-orthogonal CDMA is used for applications in which a fixed set of users continually communicate with the BS, whereas our scheme corresponds to a random subset of users in a large network communicating single bits to the BS. Second, MUD schemes for non-orthogonal CDMA require that the BS has knowledge of the individual user delays, whereas we assume—partly because of the random user activity—that user delays are unknown at the BS.

In terms of related prior work, Fletcher et al. [2] have also recently studied the problem of MUD in on–off RACs. However, the results in [2]—while similar in spirit to the ones in here—are limited by the facts that [2]: ($i$) assumes perfect synchronization among the $M$ users, which is hard to guarantee in practical settings for large $M$; ($ii$) assumes that CSIs of the individual users are available to the BS in certain cases, which is difficult—if not impossible—to justify for the case of *fading* RACs; and ($iii$) only guarantees that the probability of error $P_{err}$ at the BS goes to zero asymptotically in $M$, which does not shed light on the scaling of $P_{err}$. More recently, we have become aware of the independent and simultaneous work in [4] and [5] that also considers on–off RACs in the context of configuration in ad-hoc wireless networks. However, [4] and [5] also make a synchronization assumption similar to [2]. Finally, the work presented here also has implications in the area of sparse signal recovery, and it relates to some recent work in model selection and compressed sensing [6], [7]. We defer a detailed discussion of these implications and relationships to later parts of the paper.

We use the following notational conventions throughout the rest of the paper. We use lowercase and uppercase bold-faced letters, such as $\mathbf{x}$ and $\mathbf{X}$, to represent vectors and matrices, respectively, while we use $(\cdot)^{\mathrm{T}}$ to denote transposition of vectors and matrices. The identity matrix and the all-zeros vector are denoted by $\mathbf{I}$ and $\mathbf{0}$, respectively, and their dimensions are either given by context or explicitly shown in subscripts. The notation $\mathcal{N}(m, \sigma^2)$ signifies the Gaussian distribution with mean $m$ and standard deviation $\sigma$, $\mathsf{binary}(\pm 1/\sqrt{N}, \mathbf{I}_N)$ denotes an $N$-length Rademacher distribution in which each entry independently takes value $+1/\sqrt{N}$ or $-1/\sqrt{N}$ each with probability $1/2$, and $\mathbb{E}[\cdot]$ denotes the expectation of a random variable. We use $\Pr(\cdot)$ to denote the probability of an event and $\Pr(\cdot|\mathcal{C})$ as the probability conditioned on an event $\mathcal{C}$. The notation $\langle \mathbf{x}, \mathbf{y} \rangle$ is used to denote the inner product between vectors $\mathbf{x}$ and $\mathbf{y}$. Finally, $\log(\cdot)$ is taken as the natural logarithm throughout the paper.

The remainder of the paper is organized as follows. In Section II, we introduce our system model and accompanying assumptions. In Section III, we describe our approach to MUD for asynchronous (non-5orthogonal) CDRA and specify its performance for both arbitrary and random delays in terms of two metrics of the set of user codewords. In Section IV, we specialize the results of Section III to random binary codewords and specially constructed algebraic codewords. We finally conclude in Section V by reporting results of some numerical experiments and discussing connections of our work in the area of sparse signal recovery.

## II. SYSTEM MODEL

In this section, we formalize the problem of MUD in asynchronous on–off RACs by introducing our system model and accompanying assumptions. To begin, we assume that there are a total of $M$ users in the network that communicate with the BS using waveforms of duration $T$ and (two-sided) bandwidth $W$; in other words, the total number of temporal signal space dimensions (degrees of freedom) in the uplink are $N = TW$. In this paper, we propose that users communicate using *spread spectrum* waveforms:

$$x_i(t) = \sqrt{\mathcal{E}_i} \sum_{n=0}^{N-1} x_n^i \, g(t - nT_c), \ t \in [0, T), \quad (1)$$

where $g(t)$ is a unit-energy prototype pulse ($\int |g(t)|^2 dt = 1$), $T_c \approx \frac{1}{W}$ is the *chip duration*, $\mathcal{E}_i$ denotes the transmit power of the $i$-th user, and

$$\mathbf{x}_i = \begin{bmatrix} x_0^i & x_1^i & \ldots & x_{N-1}^i \end{bmatrix}^{\mathrm{T}}, \ i = 1, \ldots, M \quad (2)$$

is the $N$-length real-valued *codeword* of unit energy ($\|\mathbf{x}_i\|_2 = 1$) assigned to the $i$-th user.

In the context of on–off RACs, we assume that on average a total of $k$ of the $M$ users transmit 1's at time $t = 0$ (without loss of generality), resulting in the following received signal at the BS

$$y(t) = \sum_{i=1}^{M} h_i \delta_i x_i(t - \tau_i) + w(t). \quad (3)$$

Here, $h_i \in \mathbb{R}$ and $\tau_i \in \mathbb{R}_+$ are the *channel fading coefficient*[3] and the *delay*[4] associated with the $i$-th user, respectively, $w(t)$ is additive white Gaussian noise (AWGN) introduced by the receiver circuitry, and $\{\delta_i\}$ are independent 0–1 Bernoulli random variables that model the random activation of the $M$ users in the sense that $\Pr(\delta_i = 1) = k/M$. Finally, we assume that user transmissions undergo independent fading and each $h_i$ has a symmetric distribution on $\mathbb{R}$ (e.g., Rayleigh fading with $h_i$ distributed as $\mathcal{N}(0, \rho_i^2)$).

---

[3]We take fading coefficients in $\mathbb{R}$ since we are assuming real-valued codewords. Modifications for the complex-valued case are tedious but straightforward.

[4]One of the major differences between [2], [4] and the setup in here is that it is assumed in [2], [4] that $\max_{i,j}(\tau_i - \tau_j) < T_c$ whereas we do not make this assumption since it is nearly impossible to satisfy this condition for large-enough values of $M$.

May 31, 2018　　　　　　　　　　　　　　　　　　　　　　　　　　　　　　　　　　　　　　　　DRAFT

Next, we define the individual *discrete delays* $\tau'_i \in \mathbb{Z}_+$ as $\tau'_i \stackrel{def}{=} \lfloor \frac{\tau_i}{T_c} \rfloor$ and define the *maximum discrete delay* $\tau \in \mathbb{Z}_+$ in the system as an upper bound on the delays satisfying $\tau \geq \max_i \tau'_i$. It is easy to see that the received signal $y(t)$ at the BS can be sampled at the chip rate to obtain an equivalent discrete approximation

$$\mathbf{y} \approx \sum_{i=1}^{M} h_i \delta_i \sqrt{\mathcal{E}_i}\, \tilde{\mathbf{x}}_i + \mathbf{w}, \tag{4}$$

which tends to be quite accurate as long as point sampling is employed and $g(t)$ is close to being a square pulse. Here, the AWGN vector $\mathbf{w}$ is distributed as $\mathcal{N}(\mathbf{0}_{N+\tau}, \mathbf{I}_{N+\tau})$, i.e., the instantaneous received signal to noise ratio (SNR) of the active users is $\mathcal{E}_i |h_i|^2$, and the vectors $\tilde{\mathbf{x}}_i \in \mathbb{R}^{N+\tau}$ are defined as

$$\tilde{\mathbf{x}}_i = \begin{bmatrix} \mathbf{0}^{\mathrm{T}}_{\tau'_i} & \mathbf{x}^{\mathrm{T}}_i & \mathbf{0}^{\mathrm{T}}_{\tau-\tau'_i} \end{bmatrix}^{\mathrm{T}}, \; i = 1, \dots, M. \tag{5}$$

The assumptions we make here are that ($i$) the maximum delay $\tau$ is known at the BS and ($ii$) each user has knowledge of the SNR at which its transmitted signal arrives at the BS (in other words, the $i$-th user knows $|h_i|$). Both these assumptions are quite reasonable from a practical perspective; in particular, if one assumes that the BS transmits a beacon signal before the users start transmitting then the last assumption follows because of reciprocity between the downlink and uplink.

Our goal now is to specify a MUD algorithm for this asynchronous CDRA scheme that returns an estimate $\widehat{\mathcal{I}}$ of the set of active users $\mathcal{I} \stackrel{def}{=} \{i : \delta_i = 1\}$ from the $(N+\tau)$-dimensional vector $\mathbf{y}$ without knowledge of the set of delays $\{\tau'_i\}$ or the set of channel coefficients $\{h_i\}$ at the BS. Note that a benchmark for any such algorithm is synchronous, orthogonal signaling-based random access, which dictates the relationship $k \leq M \leq N$. Therefore, the primary objective of our algorithm must be to successfully manage an average number of active users that is almost linear in $N$, but also service a total number of users in the uplink that is significantly larger than $N$. In addition to this primary objective, we are also interested in specifying probability of error, $P_{err} \stackrel{def}{=} \Pr(\widehat{\mathcal{I}} \neq \mathcal{I})$, and providing a low-complexity implementation of the MUD algorithm. In the next section, we propose an algorithm that explicitly takes advantage of the random user activity in the network to successfully meet all these objectives.

## III. Multiuser Detection Using The Lasso

In this section, we propse a MUD algorithm for asynchronous CDRA that is based on the mixed-norm convex optimization program known as the lasso [8]. The lasso was first proposed in the statistics literature for linear regression in underdetermined settings. In [2], the lasso has been suggested as a potential method for MUD in *synchronous* on–off RACs. However, extending the ideas of [2] to the asynchronous case using the standard lasso formulation seems very difficult. In contrast, while the MUD algorithm proposed in this section is based on the lasso, we present a rather nonconventional usage of the lasso that is specific to the problem at hand. One of our major contributions indeed is establishing that this formulation is guaranteed to yield successful MUD with high probability. The fact that further differentiates our work from [2] is that we relate the performance of the proposed MUD algorithm for both arbitrary and random delays to two simple metrics of the set of user codewords, which enables us to construct specialized codewords for different applications. The analysis carried out in this regard might also be of independent interest to researchers working on configuration (neighbor discovery) in ad-hoc wireless networks and sensor networks. These results also have connections with the area of sparse signal recovery, as noted in Section IV-A and Section V-B.

### A. Main Results

In order to make use of the lasso for MUD in asynchronous on-off RACs, we first rewrite (4) as

$$\mathbf{y} = \underbrace{\begin{bmatrix} \tilde{\mathbf{x}}_1 & \tilde{\mathbf{x}}_2 & \ldots & \tilde{\mathbf{x}}_M \end{bmatrix}}_{\widetilde{\mathbf{X}}} \tilde{\boldsymbol{\beta}} + \mathbf{w}, \tag{6}$$

where the $i$-th entry of the vector $\tilde{\boldsymbol{\beta}} \in \mathbb{R}^M$ is described as $\tilde{\beta}_i \stackrel{def}{=} h_i \delta_i \sqrt{\mathcal{E}_i}$. While (6) appears superficially similar to the standard lasso formulation, we cannot use the lasso to obtain an estimate of the set of active users $\mathcal{I}$ from (6) since the $(N + \tau) \times M$ matrix $\widetilde{\mathbf{X}}$ in (6) is unknown due to the asynchronous nature of the problem. In order to overcome this obstacle, we first define $(N + \tau) \times (\tau + 1)$ *Toeplitz matrices* $\mathbf{X}_i$ as

$$\mathbf{X}_i = \begin{bmatrix} \mathbf{x}_i & & \mathbf{0}_\tau \\ & \ddots & \\ \mathbf{0}_\tau & & \mathbf{x}_i \end{bmatrix}, \ i = 1, \ldots, M, \tag{7}$$

and observe that we can equivalently write (6) in the form

$$\mathbf{y} = \underbrace{\begin{bmatrix} \mathbf{X}_1 & \mathbf{X}_2 & \ldots & \mathbf{X}_M \end{bmatrix} \begin{bmatrix} \boldsymbol{\beta}_1^\mathrm{T} & \boldsymbol{\beta}_2^\mathrm{T} & \ldots \boldsymbol{\beta}_M^\mathrm{T} \end{bmatrix}^\mathrm{T}}_{\mathbf{X}\boldsymbol{\beta}} + \mathbf{w}, \tag{8}$$

where $\mathbf{X}$ is now a $(N + \tau) \times M(\tau + 1)$ *known* matrix, which we term the *expanded codebook*. The vector $\boldsymbol{\beta} \in \mathbb{R}^{M(\tau+1)}$ is a concatenation of $M$ vectors, each of length $(\tau + 1)$, whose entries are given by $\beta_{i,j} = \tilde{\beta}_i 1_{\{\tau'_i = j-1\}}$, $i = 1, \ldots, M$, $j = 1, \ldots, \tau + 1$. We make use of this notation to describe the proposed lasso-based MUD algorithm for asynchronous CDRA in Algorithm 1.[5]

---

**Algorithm 1** Multiuser Detection in Asynchronous On–Off Random Access Channels Using the Lasso

**Inputs**

1) The chip-rate sampled vector $\mathbf{y}$

2) Set of $N$-dimensional codewords $\{\mathbf{x}_i\}_{i=1}^M$

3) Maximum discrete delay $\tau$ in the uplink

4) A regularization parameter $\lambda$ for the lasso

**Construct** the expanded codebook $\mathbf{X}$ described in (8) using $\{\mathbf{x}_i\}$ and $\tau$

$$\widehat{\boldsymbol{\beta}} \leftarrow \arg\min_{\mathbf{b}\in\mathbb{R}^{M(\tau+1)}} \tfrac{1}{2}\|\mathbf{y}-\mathbf{Xb}\|_2^2 + \lambda\|\mathbf{b}\|_1 \qquad \text{(LASSO)}$$

$$\widehat{\mathcal{I}} \leftarrow \left\{i : \|\widehat{\boldsymbol{\beta}}_i\|_0 > 0\right\}$$

**Return** $\widehat{\mathcal{I}}$ as an estimate of the set of active users $\mathcal{I}$

---

We next state the main results of this section, which bound the probability of error of Algorithm 1. Here we present MUD guarantees for arbitrary codebooks, parametrized by two metrics of the expanded codebook $\mathbf{X}$. The first is the *worst-case coherence* of the expanded codebook, defined by

$$\mu(\mathbf{X}) \stackrel{def}{=} \max_{(i,j)\neq(i',j')} \left|\langle \mathbf{x}_{i,j}, \mathbf{x}_{i',j'}\rangle\right| \qquad (9)$$

where $\mathbf{x}_{i,j}$ denotes the $j$-th column of the Toeplitz matrix $\mathbf{X}_i$. In words, the worst-case coherence is the largest inner product between any two codewords with arbitrary shifts. The second metric is the spectral norm of the expanded codebook: $\|\mathbf{X}\|_2 \stackrel{def}{=} \sqrt{\lambda_{max}(\mathbf{X}^{\mathrm{T}}\mathbf{X})}$.

*Theorem 1:* Suppose that users in the network become active according to independent and identically distributed (iid) Bernoulli random variables such that $\Pr(\delta_i = 1) = k/M$, and the users have transmit powers satisfying

$$\mathcal{E}_i > \frac{128\log\left(M\sqrt{\tau+1}\right)}{|h_i|^2},\ i\in\mathcal{I}. \qquad (10)$$

---

[5]Algorithm 1 acts as a hybrid between the standard lasso and the group lasso [9]. Specifically, it is clear from the problem formulation that the group lasso is ill-suited for the specified MUD problem since each of the sub-vectors $\{\boldsymbol{\beta}_i\}$ in (8) has at most one non-zero entry. On the other hand, we are only interested in detecting the active users and need not estimate their delays; hence, the group nature of the detection criterion in the definition of $\widehat{\mathcal{I}}$.

Then, with $\lambda = 2\sqrt{2\log(M\sqrt{\tau+1})}$, Algorithm 1 successfully carries out multiuser detection with $P_{err} \leq 2M^{-1}\big(2\pi \log(M\sqrt{\tau+1})\big)^{-1/2} + 5\big(M(\tau+1)\big)^{-2\log 2} + 3M^{-2\log 2}$ when

$$M \leq \frac{\exp\big((c(\tau+1)\mu(\mathbf{X}))^{-1}\big)}{\tau + 1} \quad \text{and} \quad (11)$$

$$k \leq \frac{M}{c \log\big(M(\tau+1)\big)\|\mathbf{X}\|_2^2}. \quad (12)$$

Here, the constant $c > 0$ is independent of the problem parameters.

*Remark 1:* Notice that Theorem 1 requires the transmit powers of all active users to satisfy (10). This could lead to unrealistic demands on the transmit powers of users with very small fading coefficients. There is, however, a straightforward extension of Theorem 1 that handles such situations by requiring users with small-enough fading coefficients to remain inactive. Since the required analysis in that case can be carried out by using well-known techniques for computing outage probabilities, we have chosen to forgo a detailed discussion of this issue for brevity of exposition.

The proof of this theorem is provided in Appendix A. From (11) we see that to accommodate a large number of total users $M$, we need codewords that result in an expanded codebook with a small worst-case coherence $\mu(\mathbf{X})$. Similarly, (12) shows that codewords that result in small spectral norm $\|\mathbf{X}\|_2$ allow the value of $k$ to be large. While Theorem 1 is general, it may be a bit opaque to some readers. Once applied to specific codewords in Theorems 3 and 4, however, favorable scaling relations between $k, M$ and $N$ become apparent.

Note that Theorem 1 considers recovery in the presence of an arbitrary set of delays $\{\tau_i\}$. Specifically, the result describes average-case behavior for user activity and worst-case behavior for the set of users' delays. This is desirable since fixing a probability distribution on the delays restricts the applicability of our results to only certain classes of networks. Nonetheless, considering random delays can be desirable in certain cases. To this effect, we now place a probability model on the set of delays and derive result analogous to that of Theorem 1. Explicitly, for each $i$ with $\delta_i = 1$, we consider $\tau_i$ selected uniformly at random from the set $\{0, 1, \ldots, \tau\}$. In doing so, we find that the requirement on $k$ scales more favorably with respect to the maximum delay $\tau$ when one considers this *typical-case* analysis of $\{\tau_i\}$. Note that while any probability model on the delays reduces applicability of the corresponding results to certain network settings, the uniform distribution of delays is mainly an illustrative model that is also amenable to analysis.

*Theorem 2:* Suppose that users in the network become active according to iid Bernoulli random variables such that $\Pr(\delta_i = 1) = k/M$. Further, suppose that the delays of active users $\{\tau_i : i \in \mathcal{I}\}$ are drawn uniformly at random from $\{0, 1, \ldots, \tau\}$ and the transmit powers of users satisfy (10). Then, with $\lambda = 2\sqrt{2\log(M\sqrt{\tau+1})}$, Algorithm 1 successfully carries out multiuser detection with $P_{err} \leq 2M^{-1}\big(2\pi\log(M\sqrt{\tau+1})\big)^{-1/2} + 7\big(M(\tau+1)\big)^{-2\log 2}$ when

$$M \leq \frac{\exp\big((c'(\tau+1)\mu(\mathbf{X}))^{-1}\big)}{\tau+1} \quad \text{and} \tag{13}$$

$$k \leq \frac{M(\tau+1)}{c'\log\big(M(\tau+1)\big)\|\mathbf{X}\|_2^2}. \tag{14}$$

Here, the constant $c' > 0$ is independent of the problem parameters.

The proof of this theorem is provided in Appendix B. Compared with Theorem 1, we note that the bound on the number of active users in this case scales with an additional factor of $\tau + 1$. When we specialize this theorem to random codewords in Section IV-A and a deterministic codeword construction in Section IV-B, this translates to a mere log-order dependence on $\tau$.

*B. Computational Complexity*

Theorem 1 characterizes the performance of Algorithm 1 for MUD in asynchronous on–off RACs but fails to shed any light on its computational complexity. However, the lasso is a well-studied program in the statistics literature and—thanks to its convex nature—there exist a number of extremely fast (polynomial-time) implementations of the optimization program specified in (LASSO); see, e.g., [10].

In this regard, computational complexity of the implementations of (LASSO) such as SpaRSA [10] is determined—to a large extent—by the complexity of the matrix–vector multiplications $\mathbf{Xb}$ and $\mathbf{X}^{\mathrm{T}}\mathbf{y}$. It therefore seems that Algorithm 1 increases the computational complexity of the matrix–vector multiplications from $O(NM)$, corresponding to the case of perfectly-known user delays [cf. (6)], to $O(NM(\tau+1))$. This observation, however, ignores the fact that $\mathbf{X}$ in (8) has a Toeplitz-block structure. Specifically, if we write $\mathbf{b} \in \mathbb{R}^{M(\tau+1)}$ as $\mathbf{b} = \begin{bmatrix} \mathbf{b}_1^{\mathrm{T}} & \ldots & \mathbf{b}_M^{\mathrm{T}} \end{bmatrix}^{\mathrm{T}}$ then it follows from elementary signal processing that

$$\mathbf{Xb} = \sum_{i=1}^{M} \mathcal{F}_{N+\tau}^{-1}\Big(\mathcal{F}_{N+\tau}(\mathbf{x}_i) \odot \mathcal{F}_{N+\tau}(\mathbf{b}_i)\Big), \tag{15}$$

where $\mathcal{F}_n(\cdot)$ and $\mathcal{F}_n^{-1}(\cdot)$ denote the FFT implementation of the $n$-point discrete Fourier transform (DFT) and the $n$-point inverse DFT of a sequence, respectively, while $\odot$ denotes pointwise multiplication. Similarly, if we use $(\cdot)[n_1 : n_2]$ to denote the $n_1$-th to $n_2$-th elements of a vector and $(\cdot)^-$ to denote the time-reversed version of a vector, then it follows from routine calculations that $\forall\, i = 1, \ldots, M$, we have

$$\mathbf{X}^{\mathrm{T}}\mathbf{y}[i(\tau+1) - \tau : i(\tau+1)] = \mathcal{F}_{2N+\tau-1}^{-1}\Big(\mathcal{F}_{2N+\tau-1}(\mathbf{x}_i^-) \odot \mathcal{F}_{2N+\tau-1}(\mathbf{y})\Big)[N : N+\tau].$$

It therefore follows from the complexity of the FFT that the matrix–vector multiplications $\mathbf{Xb}$ and $\mathbf{X}^{\mathrm{T}}\mathbf{y}$ in Algorithm 1 can in fact be carried out using only $O(NM\log(N+\tau))$ operations as opposed to

TABLE I

AVERAGE RECOVERY TIMES IN MATLAB FOR $N = 1023$, $M = 3072$, AND $k = 50$

| Maximum delay $\tau$ | 50 | 100 | 150 | 200 | 250 |
|---|---|---|---|---|---|
| Standard SpaRSA | 21.2s | 61.6s | 96.2s | 142.6s | 173.0s |
| FFT Augmented SpaRSA | 54.4s | 53.4s | 84.3s | 98.0s | 78.2s |

$O(NM(\tau + 1))$ operations. This suggests that the computational complexity of Algorithm 1 at worst differs by a factor of $\log(N + \tau)$ from an oracle-based scheme that has perfect knowledge of $\widetilde{\mathbf{X}}$.

This conclusion is also justified numerically from the results of several numerical experiments reported in Section V. Table I shows typical computation times of Algorithm 1 in Matlab for various values of $\tau$. The standard SpaRSA recovery is faster at low values of $\tau$ due to Matlab's optimized matrix multiplications. However, for $\tau \geq 100$, the advantage of the FFT-based implementation becomes apparent. The non-monotonicity of recovery times in the FFT augmented numerical experiments is due to the complex interaction between padding in Matlab's FFT implementation, numerical accuracy, and additional SpaRSA iterations, a detailed discussion of which is beyond the scope of this paper. Of course, for practical applications, optimizations are required beyond a Matlab implementation.

## IV. CODEWORDS FOR MULTIUSER DETECTION

In this section, we consider two sets of user codewords for asynchronous CDRA using Algorithm 1. The first is a random construction with normalized iid $\pm 1$ user codewords and the second is a deterministic construction based on cyclic codes. The deterministic construction has the advantage that user codewords can be more efficiently stored and generated. However, the random construction is more flexible with regard to codeword length and number of codewords available. Using Theorems 1 and 2, we show that (ignoring $\tau$) both sets of codewords allow the recovery of $\mathcal{I}$ when $k \lesssim N/\log M$. Furthermore, the random codewords allow $M$ to be super-polynomial in $N$ while the deterministic codewords allow $M$ to be polynomial in $N$.

### A. Random Rademacher Codewords: Guarantees

Communication theory often uses random codewords for optimality. Furthermore, random measurement matrices are frequently used in sparse signal recovery. These examples inspire us to analyze randomly generated codewords in the context of Theorems 1 and 2 for MUD in RACs.

In the following, we assign each user a codeword of length $N$ that is independently generated from a $\mathsf{binary}(\pm 1/\sqrt{N}, \mathbf{I}_N)$ distribution. We seek to quantify $\mu(\mathbf{X})$ and $\|\mathbf{X}\|_2$ of the expanded codebook to specialize Theorems 1 and 2 to these random codewords. This is accomplished in the following lemmas.

*Lemma 1:* Given any fixed $\varsigma > 0$, the expanded codebook $\mathbf{X}$ of random Rademacher codewords satisfies $\mu(\mathbf{X}) \leq \varsigma$ with probability exceeding $1 - 2M^2(\tau+1)^2 e^{-\frac{N\varsigma^2}{4}}$.

*Proof:* The proof of this lemma is a consequence of the bound on the worst-case coherence $\mu$ of random Toeplitz matrices [11, Theorem 3.5] and the Hoeffding inequality [12]. Specifically, we can write
$$\mu(\mathbf{X}) = \max\left\{\max_{j \neq j'} \left|\langle \mathbf{x}_{i,j}, \mathbf{x}_{i,j'}\rangle\right|, \max_{i \neq i'} \left|\langle \mathbf{x}_{i,j}, \mathbf{x}_{i',j'}\rangle\right|\right\}.$$
Furthermore, the proof of Theorem 3.5 in [11] implies that $\left|\langle \mathbf{x}_{i,j}, \mathbf{x}_{i,j'}\rangle\right| \leq \varsigma$ with probability exceeding $1 - 4e^{-\frac{N\varsigma^2}{4}}$ for any $j \neq j'$. Finally, since the product of two independent binary random variables is again a binary random variable, it can also be shown using the Hoeffding inequality that $\left|\langle \mathbf{x}_{i,j}, \mathbf{x}_{i',j'}\rangle\right| \leq \varsigma$ with probability exceeding $1 - 2e^{-\frac{N\varsigma^2}{2}}$ for any $i \neq i'$. It therefore follows from the union bound that $\mu(\mathbf{X}) \leq \varsigma$ with probability exceeding $1 - 2M^2(\tau+1)^2 e^{-\frac{N\varsigma^2}{4}}$. ∎

*Lemma 2:* The spectral norm of the expanded codebook $\mathbf{X}$ of random Rademacher codewords satisfies
$$\|\mathbf{X}\|_2 \leq 26\sqrt{\tau+1}\left(1 + \sqrt{\frac{M}{N}}\right) \tag{16}$$
with probability exceeding $1 - e^{-\frac{\sqrt{NM}}{8}}$.

*Proof:* We first recall that the spectral norm is invariant under column-interchange operations. Now define $\boldsymbol{\Phi} \stackrel{def}{=} \begin{bmatrix} \mathbf{x}_1 & \ldots & \mathbf{x}_M \end{bmatrix}$ and $\boldsymbol{\Psi} \stackrel{def}{=} \begin{bmatrix} \boldsymbol{\Phi}_0 & \boldsymbol{\Phi}_1 & \ldots & \boldsymbol{\Phi}_\tau \end{bmatrix}$, where each block $\boldsymbol{\Phi}_i$ is an $(N+\tau) \times M$ matrix that is constructed by prepending and appending $\boldsymbol{\Phi}$ with $i$ rows and $(\tau - i)$ rows of all zeros, respectively. It is then easy to see that $\|\mathbf{X}\|_2 = \|\boldsymbol{\Psi}\|_2$ and $\|\boldsymbol{\Phi}_0\|_2 = \cdots = \|\boldsymbol{\Phi}_\tau\|_2 = \|\boldsymbol{\Phi}\|_2$. Furthermore, we can write for any $M(\tau+1)$-dimensional vector $\mathbf{z} = \begin{bmatrix} \mathbf{z}_0^\mathrm{T} & \mathbf{z}_1^\mathrm{T} & \ldots \mathbf{z}_\tau^\mathrm{T} \end{bmatrix}^\mathrm{T}$
$$\begin{aligned}\frac{\|\boldsymbol{\Psi}\mathbf{z}\|_2}{\|\mathbf{z}\|_2} &\stackrel{(a)}{\leq} \frac{\sum_{i=0}^\tau \|\boldsymbol{\Phi}_i \mathbf{z}_i\|_2}{\|\mathbf{z}\|_2} \leq \frac{\|\boldsymbol{\Phi}\|_2 \sum_{i=0}^\tau \|\mathbf{z}_i\|_2}{\|\mathbf{z}\|_2}\\ &\stackrel{(b)}{\leq} \frac{\sqrt{\tau+1}\|\boldsymbol{\Phi}\|_2 \|\mathbf{z}\|_2}{\|\mathbf{z}\|_2} = \sqrt{\tau+1}\|\boldsymbol{\Phi}\|_2,\end{aligned} \tag{17}$$
where $(a)$ follows from the definition of $\boldsymbol{\Psi}$ and the triangle inequality, while $(b)$ follows from the Cauchy–Schwarz inequality. It therefore follows from the previous discussion and (17) that $\|\mathbf{X}\|_2 \leq \sqrt{\tau+1}\|\boldsymbol{\Phi}\|_2$.

In order to complete the proof, notice that $\boldsymbol{\Phi}$ is an $N \times M$ random matrix whose entries are independently distributed as $\mathsf{binary}(\pm 1/\sqrt{N})$. It can therefore be established, similar to [13, Proposition 2.4], that $\|\boldsymbol{\Phi}\|_2 \leq 26\left(1 + \sqrt{\frac{M}{N}}\right)$ with probability exceeding $1 - e^{-\frac{\sqrt{NM}}{8}}$. ∎

We now want to apply these lemmas to specialize Theorems 1 and 2 to the expanded codebook matrix of random codewords. We begin by noting that with $\varsigma$ of Lemma 1 chosen appropriately, the event

$$\mathcal{G}_1 = \left\{\mu(\mathbf{X}) \leq \sqrt{\frac{12\log(M(\tau+1))}{N}}\right\} \tag{18}$$

holds with probability exceeding $1 - 2\big(M(\tau+1)\big)^{-1}$. Furthermore, Lemma 2 implies that the event

$$\mathcal{G}_2 = \left\{\|\mathbf{X}\|_2 \leq 52\sqrt{\frac{M(\tau+1)}{N}}\right\} \tag{19}$$

holds with probability exceeding $1 - e^{-\frac{\sqrt{NM}}{8}}$.

Since we assume that the random codewords are assigned independently of the set of active users $\mathcal{I}$, we can substitute (18) and (19) into (11) and (12) while adding the failure probabilities $\Pr(\mathcal{G}_1^c)$ and $\Pr(\mathcal{G}_2^c)$ to the probabilities of error in Theorems 1 and 2 via the union bound. This results in the following theorem.

*Theorem 3:* Suppose that the $M$ codewords $\{\mathbf{x}_i \in \mathbb{R}^N\}_{i=1}^M$ are drawn independently from a $\mathsf{binary}(\pm 1/\sqrt{N}, \mathbf{I}_N)$ distribution. Furthermore, let $\lambda$ and $\mathcal{E}_i$ satisfy the conditions in Theorem 1 and let $M$ satisfy

$$M \leq \frac{\exp\big(c_{1,r}(\tau+1)^{-2/3}N^{1/3}\big)}{\tau+1}. \tag{20}$$

(a) For an arbitrary set of user delays, if

$$k \leq \frac{c_{2,r}N}{(\tau+1)\log(M(\tau+1))}, \tag{21}$$

then Algorithm 1 successfully carries out multiuser detection with $P_{err} \leq 2M^{-1}\big(2\pi\log(M\sqrt{\tau+1})\big)^{-1/2} + 5\big(M(\tau+1)\big)^{-2\log 2} + 3M^{-2\log 2} + 2(M(\tau+1))^{-1} + e^{\frac{\sqrt{NM}}{8}}$.

(b) For a set of user delays distributed uniformly at random, when

$$k \leq \frac{c_{3,r}N}{\log(M(\tau+1))}, \tag{22}$$

then Algorithm 1 successfully carries out multiuser detection with $P_{err} \leq 2M^{-1}\big(2\pi\log(M\sqrt{\tau+1})\big)^{-1/2} + 7\big(M(\tau+1)\big)^{-2\log 2} + 2(M(\tau+1))^{-1} + e^{\frac{\sqrt{NM}}{8}}$.

Here, the constants $c_{1,r}, c_{2,r}, c_{3,r} > 0$ are independent of the problem parameters.

*Remark 2:* It is important to note here that, instead of relying upon Theorem 1, if one were to directly analyze the MUD performance of Algorithm 1 for random codewords then it is possible to achieve the scaling $k \precsim N/\log^5(M(\tau+1))$ in the case of arbitrary delays by using the results of [7] for the "invertibility condition" in Appendix A. Specifically, the work in [7] considers random Toeplitz-block matrices in a similarly structured problem and achieves only a poly-logarithmic dependence on $\tau$ in the case of arbitrary delays. The analysis in [7], however, is not extendable to arbitrary Toeplitz-block matrices and further the proof techniques used in there introduce some complications related to noise folding. In contrast, our focus in here is to provide conditions applicable to arbitrary codewords via the metrics $\|\mathbf{X}\|_2$ and $\mu(\mathbf{X})$, and our results for random matrices/codewords are primarily meant to be a demonstration of our more general results. Nonetheless, we believe that [7] provides unique insights for Algorithm 1 in the case of random codewords and arbitrary delays.

*B. Deterministic Codewords: Construction and Guarantees*

Though random codewords allow our proposed scheme to service a large number of users (with respect to $N$), deterministic codewords can have significant advantages. In particular, they tend to be much easier to generate and store. We will consider one such codeword construction in this section.

Our deterministic construction uses codewords derived from algebraic error correcting codes. In particular, we consider a cyclic code for which the codebook is closed under circular shifts of codewords. We use a cyclic code since we use cyclic shifts as approximations of delayed user codewords. As such, the full cyclic code is closely related to the expanded codebook matrix $\mathbf{X}$ which contains the delayed user codewords. To construct this relationship, we will select a subset of our cyclic code for assignment to users. In order to remove ambiguity when discussing both the full cyclic code and this subset, we will call the complete cyclic code the *ambient code*, while the subset assigned to users will be called the *user codebook* (i.e., the user codebook is the set $\{\mathbf{x}_i\}$).

Our construction is parametrized by two positive integers, $m$ and $2 \leq t < m/2$. We will operate in the Galois finite field of size $2^m$, which we denote as $\mathrm{GF}(2^m)$. Our code is constructed via the trace function $\mathrm{Tr} : \mathrm{GF}(2^m) \to \mathrm{GF}(2)$ ([14, Ch. 4.8]) defined by

$$\mathrm{Tr}(a) = a + a^2 + \cdots + a^{2^{m-1}} = \sum_{j=0}^{m-1} a^{2^j}.$$

Taking $z$ as a primitive element of $\mathrm{GF}(2^m)$ we define the $j^{\text{th}}$ element of a codeword in the ambient code as

$$C_\alpha^j = \frac{1}{\sqrt{2^m - 1}} (-1)^{\mathrm{Tr}\left[\alpha_0 z^j + \sum_{i=1}^{t} \alpha_j z^{j(2^i+1)}\right]}, \qquad j = 0, 1, 2, \ldots, 2^m - 2 \qquad (23)$$

where the vector $\alpha = \begin{bmatrix} \alpha_0 & \alpha_1 & \cdots & \alpha_t \end{bmatrix}$ with $t + 1$ elements in $\mathrm{GF}(2^m)$ indexes the codeword.

Since $z$ is primitive, $\{z^j : j = 0, 1, \ldots, 2^m - 2\}$ is simply the set of all non-zero elements in $\mathrm{GF}(2^m)$, which we denote $\mathrm{GF}(2^m)^*$. Thus, we can equivalently enumerate the elements by $x \in \mathrm{GF}(2^m)^*$ as

$$C_\alpha^{(x)} = \frac{1}{\sqrt{2^m - 1}} (-1)^{\mathrm{Tr}\left[\alpha_0 x + \sum_{i=1}^{t} \alpha_i x^{2^i+1}\right]}, \qquad x \in \mathrm{GF}(2^m)^*. \qquad (24)$$

The above construction produces codewords of length $2^m - 1$ and, since each of $\alpha_i, i = 0, \ldots, t$ can be any value in $\text{GF}(2^m)$, there are $2^{m(t+1)}$ codewords in the ambient code.

We use a subset of the ambient code as the user codebook. We will require two conditions on the selected subset. The first condition restricts us to a subset where no codeword in the subset is a cyclic shift of another. Such a restriction is necessary since, in bounding $\mu(\mathbf{X})$, we will link cyclic shifts with different user delays. We will call such a set a *cyclic restricted subset*. There are many ways to create a cyclic restricted subset. Consider that, under the element enumeration of (23), a codeword and its cyclic shift by $T$ elements are related as $C_\alpha^{j+T} = C_{\alpha'}^j$ where $\alpha = [\alpha_0, \ldots, \alpha_t]$ and $\alpha' = [z^T \alpha_0, z^{3T} \alpha_1, \ldots, z^{(2^t+1)T} \alpha_t]$. If, for example, we required that all codewords in our subset had $\alpha_0 = c$ for $c \in \text{GF}(2^m)^*$, no codewords would be the shift of another. Explicitly, the codewords indexed by $\{\alpha \in \text{GF}(2^m)^t : \alpha_0 = c\}$ form a cyclic restricted subset for any $c \neq 0 \in \text{GF}(2^m)$. Since we need only restrict a single entry in the vector $\alpha$, enumerating over the remaining entries allows us to have a cyclic restricted subset of size $2^{mt}$ from the full $2^{m(t+1)}$ ambient code. We may choose to use a smaller set, affording flexibility in choosing the value of $M$, and the set would remain a cyclic restricted subset.

The second condition on selecting the user codebook as a subset of the ambient code is used to ensure we can appropriately bound $\|\mathbf{X}\|_2$. As above, our condition will be on the set of vectors $\alpha$ which enumerate the codewords in the user codebook. To describe the condition we first define a *wildcard index* of the user codebook. We call $w$ a wildcard index of the user codebook if, for each vector $\alpha = [\alpha_0, \ldots, \alpha_w, \ldots, \alpha_t]$ that indexes a user codeword, the vector $[\alpha_0, \ldots, c, \ldots, \alpha_t]$ (i.e., the vector $\alpha$ with $\alpha_w$ replaced by $c$) also indexes a user codeword for each $c \in \text{GF}(2^m)$. We require that the user codebook have a wildcard index $w$ such that $2^w + 1$ does not divide $2^m - 1$ (denoted $2^w + 1 \nmid 2^m - 1$). A consequence of requiring the existence of a wildcard index is that the user codebook must be a multiple of $2^m$ in size.

To summarize our construction, we assign to users codewords of the form (23). We require that the user codebook satisfy two conditions. The first is that it forms a cyclic restricted subset of the ambient code of all possible codewords. The second is that the user codebook contain a wildcard index $w$ with $2^w + 1 \nmid 2^m - 1$. This construction allows us to have $N = 2^m - 1$ while $M$ may be a multiple of $2^m$ up to $2^{mt}$.

We now seek to apply Theorems 1 and 2 to this construction which requires us to bound the two metrics $\mu(\mathbf{X})$ and $\|\mathbf{X}\|_2$. We consider each of the two metrics in turn. The metric $\mu(\mathbf{X})$ bounds inner products between any two shifted codewords in the user codebook. As discussed earlier, we will be relating the set of shifted codewords to the ambient code. As a result, our first goal is to bound the inner product of any two codewords in the ambient code. The bound can be obtained easily by exploiting properties of the ambient code. By the linearity of the trace [14, p.116], we find that the element-wise product of two codewords satisfies $C_\alpha^j C_{\alpha'}^j = C_{\alpha+\alpha'}^j$.[6] As a result, the inner product between two non-identical codewords is simply the sum of the entries of a different codeword. This leads to an equivalent goal of bounding the sum of an arbitrary non-trivial codeword. That is, allowing $\{\alpha_i \in \mathrm{GF}(2^m)\}_{i=0}^t$ to be arbitrary with $\alpha \neq 0$, we attempt to bound the sum

$$S = \sum_{x \in \mathrm{GF}(2^m)^*} (-1)^{\mathrm{Tr}\left[\alpha_0 x + \sum_{i=1}^t \alpha_i x^{2^i+1}\right]}. \tag{25}$$

*Lemma 3:* The sum given in (25) satisfies $|S| \leq 2^{\frac{m}{2}+t+1/2}$ for any codeword.

This lemma is proved in Appendix C. We leverage this result to provide a bound on $\mu(\mathbf{X})$.

*Lemma 4:* Let the user codebook $\{\mathbf{x}_i\}$ be a cyclic restricted subset of the ambient code defined by (23). Then the worst-case coherence is bounded by

$$\mu(\mathbf{X}) \leq \frac{2^{t+1/2+m/2} + \tau}{2^m - 1}.$$

*Proof:* We are interested in bounding $|\langle \mathbf{x}_{i,j}, \mathbf{x}_{i',j'} \rangle|$. We will do so by relating this inner product to one in the ambient code. Using the fact that $\mathbf{x}_{i',j'}$ is of the form (5), we can replace vectors of zeros above and below the codeword with shifted copies of $\mathbf{x}_{i'}$ in order to make the periodic vector $\hat{\mathbf{x}}_{i',j'}$ with period $N$ on its $N+\tau$ length; we call $\hat{\mathbf{x}}_{i',j'}$ the *periodic extension* of $\mathbf{x}_{i',j'}$. Let $\Delta = \mathbf{x}_{i',j'} - \hat{\mathbf{x}}_{i',j'}$ be its difference from the original. By the triangle inequality,

$$|\langle \mathbf{x}_{i,j}, \mathbf{x}_{i',j'} \rangle| \leq |\langle \mathbf{x}_{i,j}, \hat{\mathbf{x}}_{i',j'} \rangle| + |\langle \mathbf{x}_{i,j}, \Delta \rangle|.$$

The periodic extension has converted the shift $j'$ into a cyclic shift on the support of $\mathbf{x}_{i,j}$. Furthermore, since the two unshifted codewords $\mathbf{x}_i$ and $\mathbf{x}_{i'}$ come from a cyclic restricted subset, they are guaranteed to be different on the support. Thus, for the first term we use the bound of Lemma 3 divided by $2^m - 1$, thereby accounting for the normalization of the users' codewords. The second term is bounded by the fact that the support of $\Delta$ overlaps with that of $\mathbf{x}_{i,j}$ with at most $\tau$ elements of value $\pm 1/\sqrt{2^m - 1}$. ∎

Having bounded $\mu(\mathbf{X})$, we now turn to the second metric $\|\mathbf{X}\|_2$ and bound it using the following lemma.

*Lemma 5:* Let the user codebook of cardinality $M$ selected from the ambient code have a wildcard index $w$ such that $2^w + 1 \nmid 2^m - 1$. Then the spectral norm of the expanded codebook $\mathbf{X}$ is bounded by $\|\mathbf{X}\|_2 \leq \sqrt{\frac{M}{2^m - 1}(\tau + 1)}$.

---

[6]This is a simple reformulation of the fact that the non-exponentiated version of the code in (23) is linear.

*Proof:* We begin the proof similarly to Lemma 2. Let $\boldsymbol{\Phi}$ be an $N \times M$ matrix of the user codewords and recall from Lemma 2 that $\|\mathbf{X}\|_2 \leq \sqrt{\tau+1}\|\boldsymbol{\Phi}\|_2$. We will now show that the rows of $\boldsymbol{\Phi}$ are orthogonal, such that $\boldsymbol{\Phi}\boldsymbol{\Phi}^{\mathrm{T}} = \frac{M}{2^m-1}\mathbf{I}$, which is sufficient for the proof since $\lambda_{\max}(\boldsymbol{\Phi}\boldsymbol{\Phi}^{\mathrm{T}}) = \lambda_{\max}(\boldsymbol{\Phi}^{\mathrm{T}}\boldsymbol{\Phi})$.

Using (24), the inner product between two rows, indexed by $x$ and $y$ in $\mathrm{GF}(2^m)$, is the following sum over the vectors $\alpha$ indexing the user codebook:

$$\frac{1}{2^m-1}\sum_{\alpha} C_\alpha^{(x)} C_\alpha^{(y)} = \frac{1}{2^m-1}\sum_{\alpha} (-1)^{\mathrm{Tr}\left[\alpha_0(x+y)+\sum_{i=1}^t \alpha_i(x^{2^i+1}+y^{2^i+1})\right]} \tag{26a}$$

$$= \frac{1}{2^m-1}\sum_{\alpha_w \in \mathrm{GF}(2^m)} (-1)^{\mathrm{Tr}\left[\alpha_w(x^{2^w+1}+y^{2^w+1})\right]} \sum_{\alpha\setminus\{\alpha_w\}} (-1)^{\mathrm{Tr}\left[\alpha_0(x+y)+\sum_{\substack{i=1\\i\neq w}}^t \alpha_i(x^{2^i+1}+y^{2^i+1})\right]}, \tag{26b}$$

where in (26b) we have separated the wildcard index element $\alpha_w$, which takes every value in $\mathrm{GF}(2^m)$, into a separate sum. For all $a \in \mathrm{GF}(2^m)$ we have $a + a = 0$. Thus, from the sum in (26a), when $x = y$ each term takes unit value and their sum equals $M$, the number of $\alpha$. In the case when $x \neq y$, we examine (26b). By our wildcard index condition we are guaranteed that $x^{2^w+1} \neq y^{2^w+1}$ so that $\alpha_w$ has a non-zero coefficient in the trace. By Proposition 5(a) in Appendix C, precisely half the terms of the sum over $\alpha_w$ are $-1$ and thus the whole sum evaluates to 0. Therefore, the inner product between the two rows evaluates to $M/(2^m-1)$ when $x = y$ and to 0 otherwise. This completes the proof of the lemma. ∎

Having bound both $\mu(\mathbf{X})$ and $\|\mathbf{X}\|_2$, we are able to apply Theorems 1 and 2 to this deterministic construction and give the following recovery guarantees.

*Theorem 4:* Let $m$ and $t$ be positive integers and let $N = 2^m - 1$. Suppose that the $M$ codewords $\{\mathbf{x}_i \in \mathbb{R}^N\}_{i=1}^M$ are chosen from the code defined by (23) such that they form a cyclic restricted subset and have a wildcard index $w$ with $2^w + 1 \nmid 2^m - 1$. Furthermore, let $\lambda$ and $\{\mathcal{E}_i\}$ satisfy the conditions in Theorem 1 and let $M$ satisfy,

$$M \leq \frac{\exp\left(c_{1,d}\frac{2^m-1}{(\tau+1)(2^{t+1/2+m/2}+\tau)}\right)}{\tau+1}. \tag{27}$$

(a) For an arbitrary set of user delays, if

$$k \leq \frac{c_{2,d}(2^m-1)}{(\tau+1)\log(M(\tau+1))}, \tag{28}$$

then Algorithm 1 successfully carries out multiuser detection with $P_{err} \leq 2M^{-1}\bigl(2\pi\log(M\sqrt{\tau+1})\bigr)^{-1/2} + 5\bigl(M(\tau+1)\bigr)^{-2\log 2} + 3M^{-2\log 2}$.

(b) For a set of user delays distributed uniformly at random, if

$$k \leq \frac{c_{3,d}(2^m - 1)}{\log\left(M(\tau + 1)\right)},\tag{29}$$

then Algorithm 1 successfully carries out multiuser detection with $P_{err} \leq 2M^{-1}\bigl(2\pi \log(M\sqrt{\tau+1})\bigr)^{-1/2} + 7\bigl(M(\tau+1)\bigr)^{-2\log 2}$.

Here, the constants $c_{1,d}, c_{2,d}, c_{3,d} > 0$ are independent of the problem parameters.

It is important to note here that although (27) appears to allow a super-polynomial number of users $M$, our construction restricts us to at most $2^{mt}$ codewords to be assigned to users. For small values of $t$, this restriction on $M$ dominates the one in (27). However, as $t$ approaches $\frac{m-1}{2}$, (27) becomes the relevant bound on $M$. In general, comparing our deterministic construction of codewords to the randomly generated ones in Section IV-A, we find that the proposed deterministic codewords have advantages in storage and generation while randomly generated codes have the advantages that $N$ is arbitrary and that $M$ can be super-polynomial in $N$.

## V. NUMERICAL RESULTS AND DISCUSSION

### A. Monte Carlo Experiments

To verify and illustrate the results presented in this paper for MUD in asynchronous RACs, we make use of Monte Carlo trials. Our numerical experiments assume a total of $M = 3072$ users communicating to the BS using codewords of length $N = 1023$. We report the MUD results for both the random codewords of Section IV-A and the deterministic construction of Section IV-B. For the deterministic construction, a code generated with $m = 10$ and $t = 2$ is used with a subset of the ambient code of size $M = 3(2^m)$ assigned to users. Random user activity is generated using independent 0–1 Bernoulli random variables $\{\delta_i\}$ such that $\Pr(\delta_i = 1) = k/M$ for a given $k$. Furthermore, for a given maximum delay $\tau$, the individual user delays $\{\tau_i\}$ are generated once at random for each experiment and then fixed for the remainder of the experiment. The implementation of Algorithm 1 uses the SpaRSA package [10] in order to solve (LASSO) and includes the modifications described in Section III-B for speeding up the matrix–vector multiplications $\mathbf{Xb}$ and $\mathbf{X}^T\mathbf{y}$. In all the numerical plots, results for random codewords are displayed using solid lines while those for deterministic codewords are displayed using dashed lines.

The numerical experiments correspond to the ability of the MUD scheme proposed in Algorithm 1 to correctly recover the active user set $\mathcal{I}$ for varying values of the average number of active users $k$ and maximum delay $\tau$. The results of these experiments are reported in Figure 1, which shows that when $k$ is below a certain threshold, $\mathcal{I}$ is exactly recovered (i.e., $\widehat{\mathcal{I}} = \mathcal{I}$) in the vast majority of Monte Carlo trials. Beyond the threshold of $k \approx 50$, the fraction of Monte Carlo trials in error quickly approaches one. Figure 1 also shows that codewords generated as described in Section IV-B perform nearly identically in performance to those randomly generated.

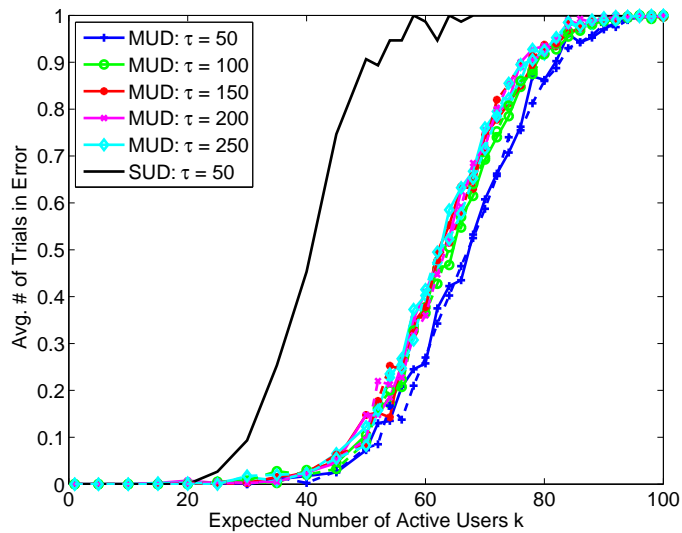

Fig. 1. User support recovery error rate as a function of the expected number of active users $k$.

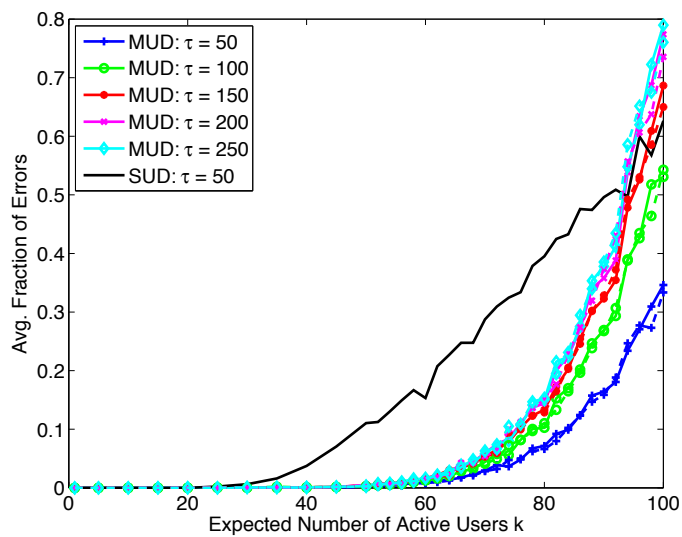

Fig. 2. Normalized per user error as a function of the expected number of active users $k$.

In order to compare our MUD results with some of the traditional SUD approaches, we have included numerical results corresponding to the performance of a matched filter receiver for the case of random codewords. We assume that the SUD receiver has access to the outputs of the matched filters for all the $M(\tau + 1)$ user codewords and shifts as well as an oracle knowledge of $|\mathcal{I}|$ (which is more specific than knowledge of $k = \mathbb{E}|\mathcal{I}|$). Consequently, the receiver declares the users corresponding to the $|\mathcal{I}|$ largest matched filter responses to be active. Note that, in general, any practical SUD receiver that detects using a fixed threshold for the matched filter responses is expected to perform worse than this oracle-like SUD. Despite this, we find that our proposed MUD algorithm significantly outperforms the traditional SUD receiver based on matched filtering ideas.

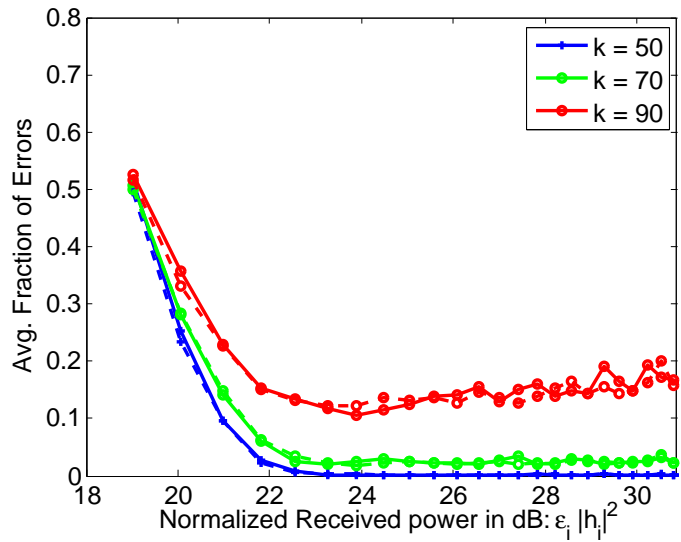

Fig. 3. Normalized per user error as a function of the received user power with $\tau = 50$.

Note that the results in Figure 1 are reminiscent of Theorem 2 and the related sections of Theorems 3 and 4, rather than Theorem 1. This is because results for the worst-case algorithmic performance are difficult to verify experimentally. While guarantees for arbitrary user delays are desirable, verifying this numerically would require generating all $(\tau + 1)^k$ possible combinations of $\{\tau_i\}$ in each Monte Carlo trial. The worst-case analyses of (21) and (28) suggest that the threshold should be inversely proportional to $(\tau + 1)$. On the other hand, Figure 1 does not exhibit this behavior since our numerical experiments correspond to a random generation of the delays $\{\tau_i\}$. Rather, they show that the recovery threshold of $k$ for a typical set of $\{\tau_i\}$ is not a strong function of $\tau$. This corresponds with the results for randomly distributed $\{\tau_i\}$ in (22) and (29).

The recovery metric in Figure 1 matches that of the theorems and declares a trial to be in error when $\widehat{\mathcal{I}} \neq \mathcal{I}$. However, it is also useful in many cases to consider how far the estimate $\widehat{\mathcal{I}}$ is from the correct set. Therefore, in Figure 2 we use the performance metric of *average fraction of detection errors*, which corresponds to $\frac{|(\mathcal{I}\setminus\widehat{\mathcal{I}})\cup(\widehat{\mathcal{I}}\setminus\mathcal{I})|}{k}$ and describes the number of errors in the estimated set of active users as a fraction of the average number of active users. With this metric, we see that Algorithm 1 fails gracefully as $k$ increases.

We also use the recovery metric of average fraction of detection errors to describe the power requirements of the active users in Figure 3. This figure shows that the power requirement, $\mathcal{E}_i$ described in Theorems 1 and 2, is overly restrictive. Specifically, the rightmost point of the horizontal axis at $\mathcal{E}_i|h_i|^2 = 31$dB provides the reference point as the power seen at the receiver as required by (10). The figure shows much less power is needed for recovering $\mathcal{I}$. It also shows that the required power is not a function of $k$ which is exactly in line with the results of our theory.

## B. Discussion

In order to place our results in context, we note that $k \lesssim N/\log M$ scaling has also been suggested in [2] for the case of MUD in synchronous on–off RACs using the lasso and random Gaussian codewords. Here, however, we provide non-asymptotic results for the more general asynchronous case, in contrast to the asymptotic results in [2]. Furthermore, we provide guarantees that can be applied to arbitrary user codewords. For the codewords studied in Theorems 3 and 4 we have established that the MUD scheme for asynchronous on–off RACs has the ability to achieve roughly the same (non-asymptotic) scaling of the system parameters $k, N$, and $M$ as that reported in [2] for the ideal case of synchronous channels.

With regard to the deterministic codewords introduced in Section IV-B, our construction is representative of a larger class of deterministic matrices derived from cyclic codes. We consider a particular cyclic code where the codewords are obtained by evaluating quadratic forms at elements of the field $\text{GF}(2^m)$. The worst case coherence $\mu(\mathbf{X})$ of the expanded codebook matrix is determined by the minimum weight of the code and we bound this quantity by elementary methods in Appendix C. We note that Yu and Gong [15] have calculated the exact weight distribution of a very similar code using more sophisticated methods from symplectic geometry.

Beyond the application to MUD for RACs, our results can also be related to work on model selection. Most directly, our work builds on the model selection theory of Candès and Plan [6] for the lasso. As in this paper, [6] provides guarantees for lasso that are based on worst-case coherence $\mu(\mathbf{X})$ and spectral norm $\|\mathbf{X}\|_2$. However, a key assumption in [6] requires that the vector $\beta$ be "generic" in the sense that its support is uniform over its $(\tau+1)M$ elements. In this paper, however, we assume a much different model: the support of $\beta$ is uniformly random over blocks of elements. In this light, the work here is related to recent work on block-sparse signals such as [16] which considers block-sparse signal recovery using a variant of orthogonal matching pursuit as opposed to the lasso. However, as work in the context of signal recovery rather than model selection, the work in [16] is not directly concerned with estimating $\mathcal{I}$ and cannot be applied to the MUD problem in RACs.

As a study of sparse signal recovery using a structured measurement matrix, this work relates to that of Romberg and Neelamani [7]. Though [7] considers a different application and is concerned with signal recovery rather than estimating $\mathcal{I}$, it studies Toeplitz-block matrices that are similar in structure to $\mathbf{X}$. The approach in [7], however, differs from ours since they provide recovery guarantees based on the *restricted isometry property* (RIP) of the matrix. By working with the RIP, the analysis is particular to randomly generated Toeplitz columns. In contrast, here we provide guarantees for any matrix $\mathbf{X}$ with sufficiently small $\mu(\mathbf{X})$ and $\|\mathbf{X}\|_2$. Subsequently, we give both randomly generated and deterministic codeword designs satisfying the requirements. Furthermore, our work provides support set recovery guarantees—in the spirit of model selection [17]—rather than bounds on recovered signal error guaranteed by RIP.

Finally, in terms of the application of our theory in the real-world, we note that Theorems 1–4 provide non-asymptotic bounds on $k$ and $M$ that guarantee recovery of the set of active users. However, we have not shown that these bounds are tight. Indeed, numerical experiments show that the bounds are somewhat loose in practice. Nonetheless, the theory provides useful scaling relationships with the metrics $\mu(\mathbf{X})$ and $\|\mathbf{X}\|_2$ which, as we have demonstrated, can guide non-orthogonal codeword designs in practical systems.

We conclude this section by pointing out three key directions of future work in the context of random access within asynchronous network settings. One of these directions involves modifying Algorithm 1 to allow for a small fraction of missed detections at the expense of reducing the fraction of false positives. The second direction involves investigating tight converses of Theorems 1 and 2 in terms of $k$, $N$, and $M$. The last direction involves extending Theorems 1 and 2 under the assumption of multipath in the uplink. Given the structured nature of the problem discussed in here, all three of these directions present some unique analytical challenges and we expect to address those challenges in a sequel to this work.

## VI. CONCLUSION

In this paper, we described a novel scheme for MUD in RACs that allows for the user codewords to be received asynchronously at the receiver. We leveraged and generalized sparse signal theory to provide recovery guarantees for a lasso-based algorithm to find the set of active users. While our results are general and applicable to arbitrary sets of codewords, we specialized them to two specific sets of codewords, random binary codewords and specially constructed algebraic codewords.

The implications of the scaling behavior outlined in the pairs of inequalities in Theorems 3 and 4 are quite positive in the important special case of fixed-bandwidth spread spectrum waveforms and a BS serving a bounded geographic region. Specifically, they signify that—for any fixed number of temporal signal space dimensions $N$ and maximum delay $\tau$ in the system—the proposed MUD scheme can accommodate $M \lesssim \exp(O(N^{1/3}))$ total users in the case of random signaling and $M$ polynomial in $N$ when using our algebraic code design. Both sets of codewords allow $k \lesssim N/\log M$ active users in the system. This is a significant improvement over the $k \leq M \lesssim N$ scaling suggested by the use of classical matched filtering-based approaches to MUD employing orthogonal signaling.

## VII. Acknowledgments

The authors would like to thank the anonymous reviewers for their helpful comments. In particular, valuable feedback from one of the reviewers motivated the random user delay analysis of Theorem 2 and the connections to [7] for random codebooks in this regime.

This work was supported in part by NSF under grants CNS-1011962 and DMS-0914892, by ONR under grant N00014-08-1-1110, and by AFOSR under grants FA9550-09-1-0643 and FA9550-09-1- 0551. MFD was also supported by NSF Supplemental Funding DMS-0439872 to UCLA-IPAM, P.I. R. Caflisch.

## Appendix A
### Proof of the Main Result: Arbitrary Delays

In this appendix, we provide a proof of Theorem 1. Before proceeding further, however, let us develop some notation to facilitate the forthcoming analysis. Throughout this appendix, we use $\mathbf{X}_\mathcal{B}$ to denote the *block subdictionary* of $\mathbf{X}$ obtained by collecting the Toeplitz blocks of $\mathbf{X}$ corresponding to the indices of the active users: $\mathbf{X}_\mathcal{B} \stackrel{def}{=} [\mathbf{X}_i : i \in \mathcal{I}]$. In addition, we use $\mathbf{X}_\mathcal{S}$ to denote the $(N+\tau) \times |\mathcal{I}|$ submatrix obtained by collecting the columns of $\mathbf{X}$ corresponding to the nonzero entries of $\boldsymbol{\beta}$, while we use $\boldsymbol{\beta}_\mathcal{S}$ to denote the $|\mathcal{I}|$-dimensional vector comprising of the nonzero entries of $\boldsymbol{\beta}$. Finally, we use $\operatorname{sgn}(\cdot)$ for elementwise *signum* function: $\operatorname{sgn}(z) \stackrel{def}{=} z/|z|$ for any $z \in \mathbb{R}$.

The basic idea behind the proof of Theorem 1 follows from the proof of [6, Theorem 1.3]. Specifically, using $\mathcal{S} \subset \{1, \ldots, M(\tau+1)\}$ to denote the set of the locations of the nonzero entries of $\boldsymbol{\beta}$, we have from [6, Lemma 3.4] that the lasso solution $\widehat{\boldsymbol{\beta}} \stackrel{def}{=} \boldsymbol{\beta} + \mathbf{h}$ satisfies $\mathbf{h}_{\mathcal{S}^c} = 0$ and

$$\mathbf{h}_\mathcal{S} = (\mathbf{X}_\mathcal{S}^T \mathbf{X}_\mathcal{S})^{-1} [\mathbf{X}_\mathcal{S}^T \mathbf{w} - \lambda \operatorname{sgn}(\boldsymbol{\beta}_\mathcal{S})] \tag{30}$$

if $\min_{i \in \mathcal{S}} |\beta_i| > 4\lambda$ and the following five conditions are met:

- $\mathcal{C}_1$ – Invertibility condition: $\|(\mathbf{X}_\mathcal{S}^T\mathbf{X}_\mathcal{S})^{-1}\|_2 \leq 2$.
- $\mathcal{C}_2$ – Noise stability: $\|(\mathbf{X}_\mathcal{S}^T\mathbf{X}_\mathcal{S})^{-1}\mathbf{X}_\mathcal{S}^T\mathbf{w}\|_\infty \leq \lambda$.
- $\mathcal{C}_3$ – Complementary noise stability: $\|\mathbf{X}_{\mathcal{S}^c}^T(\mathbf{I} - \mathbf{X}_\mathcal{S}(\mathbf{X}_\mathcal{S}^T\mathbf{X}_\mathcal{S})^{-1}\mathbf{X}_\mathcal{S}^T)\mathbf{w}\|_\infty \leq \frac{\lambda}{\sqrt{2}}$.
- $\mathcal{C}_4$ – Size condition: $\|(\mathbf{X}_\mathcal{S}^T\mathbf{X}_\mathcal{S})^{-1}\mathrm{sgn}(\boldsymbol{\beta}_\mathcal{S})\|_\infty \leq 3$.
- $\mathcal{C}_5$ – Complementary size condition: $\|\mathbf{X}_{\mathcal{S}^c}^T\mathbf{X}_\mathcal{S}(\mathbf{X}_\mathcal{S}^T\mathbf{X}_\mathcal{S})^{-1}\mathrm{sgn}(\boldsymbol{\beta}_\mathcal{S})\|_\infty \leq \frac{1}{4}$.

Furthermore, it trivially follows in this case that the set of non-zero elements of $\widehat{\boldsymbol{\beta}}$ is $\mathcal{S}$, which guarantees that $\widehat{\mathcal{I}} = \mathcal{I}$. Our goal then is to consider the probability of each one of these conditions *not being met* under the assumptions of Theorem 1 and the proof of the theorem would then simply follow from the union bound.

## A. Invertibility Condition

In order to establish the invertibility condition, we will make use of the following proposition from [18].

*Proposition 1 ([18]):* Fix $q = 2\log(M(\tau+1))$ and define the block coherence

$$\mu_B(\mathbf{X}) \stackrel{def}{=} \max_{1 \leq i,j \leq M} \|\mathbf{X}_i^T\mathbf{X}_j - 1_{\{i=j\}}\mathbf{I}\|_2. \tag{31}$$

Then, for $\mathbb{E}_q Z \stackrel{def}{=} [\mathbb{E}|Z|^q]^{1/q}$ and $\delta \stackrel{def}{=} k/M$, we have the following bound

$$\mathbb{E}_q\|\mathbf{X}_\mathcal{B}^T\mathbf{X}_\mathcal{B} - \mathbf{I}\|_2 \leq 20\mu_B(\mathbf{X})\log(M(\tau+1)) + \delta\|\mathbf{X}\|_2^2 + 9\sqrt{\delta\log(M(\tau+1))(1+\tau\mu(\mathbf{X}))}\|\mathbf{X}\|_2. \tag{32}$$

We would like to bound (32) via bounds on $\mu_B(\mathbf{X})$, $\mu(\mathbf{X})$ and $\|\mathbf{X}\|_2$. First, we can use the linear algebra fact $\|\cdot\|_2 \leq \sqrt{\|\cdot\|_1\|\cdot\|_\infty}$ [19] on (31) to show that $\mu_B(\mathbf{X}) \leq (\tau+1)\mu(\mathbf{X})$. Thus, we can rearrange the inequalities of (11) and (12) to obtain

$$\mu(\mathbf{X}) \leq \frac{1}{c(\tau+1)\log(M(\tau+1))}, \tag{33}$$

$$\|\mathbf{X}\|_2^2 \leq \frac{1}{c\delta\log(M(\tau+1))}, \quad \text{and} \tag{34}$$

$$\mu_B(\mathbf{X}) \leq \frac{1}{c\log(M(\tau+1))}. \tag{35}$$

Substituting these inequalities into (32) and choosing $c$ appropriately large yields $\mathbb{E}_q\|\mathbf{X}_\mathcal{B}^T\mathbf{X}_\mathcal{B} - \mathbf{I}\|_2 < \frac{1}{4}$.

Finally, notice that $\mathbf{X}_\mathcal{S}$ is a submatrix of $\mathbf{X}_\mathcal{B}$ and therefore we trivially have $\|\mathbf{X}_\mathcal{S}^T\mathbf{X}_\mathcal{S} - \mathbf{I}\|_2 \leq \|\mathbf{X}_\mathcal{B}^T\mathbf{X}_\mathcal{B} - \mathbf{I}\|_2$. It can then be easily seen from the Markov inequality that

$$\Pr(\|\mathbf{X}_\mathcal{S}^T\mathbf{X}_\mathcal{S} - \mathbf{I}\|_2 > 1/2) \leq 2^q(\mathbb{E}_q\|\mathbf{X}_\mathcal{B}^T\mathbf{X}_\mathcal{B} - \mathbf{I}\|_2)^q$$

$$\stackrel{(a)}{\leq} (M(\tau+1))^{-2\log 2} \tag{36}$$

where (a) follows from the fact that $\mathbb{E}_q \|\mathbf{X}_\mathcal{B}^T \mathbf{X}_\mathcal{B} - \mathbf{I}\|_2 < \frac{1}{4}$. We have now established that $\|\mathbf{X}_\mathcal{S}^T \mathbf{X}_\mathcal{S}\|_2 \in (1/2, 3/2)$ with high probability, which implies that

$$\Pr(\mathcal{C}_1^c) \leq \left(M(\tau+1)\right)^{-2\log 2}. \tag{37}$$

### B. Noise Stability

In order to establish the noise-stability condition, we first condition on $\mathcal{C}_1$ (the invertibility condition). Next, we denote the $j$-th column of $\mathbf{X}_\mathcal{S}(\mathbf{X}_\mathcal{S}^T \mathbf{X}_\mathcal{S})^{-1}$ by $\mathbf{z}_j$ and note that

$$\|(\mathbf{X}_\mathcal{S}^T \mathbf{X}_\mathcal{S})^{-1} \mathbf{X}_\mathcal{S}^T \mathbf{w}\|_\infty = \max_{1 \leq j \leq |\mathcal{S}|} |\langle \mathbf{z}_j, \mathbf{w} \rangle|. \tag{38}$$

Furthermore, since the noise vector $\mathbf{w}$ is distributed as $\mathcal{N}(\mathbf{0}, \mathbf{I})$, we also have that $\langle \mathbf{z}_j, \mathbf{w} \rangle \sim \mathcal{N}(0, \|\mathbf{z}_j\|_2^2)$. Finally, note that conditioned on $\mathcal{C}_1$, we have the upper bound

$$\|\mathbf{z}_j\|_2 \leq \|\mathbf{X}_\mathcal{S}(\mathbf{X}_\mathcal{S}^T \mathbf{X}_\mathcal{S})^{-1}\|_2 \leq \sqrt{2}.$$

where the second inequality can be seen by considering the singular value decomposition of $\mathbf{X}_\mathcal{S}$ along with the bound on the singular values from $\mathcal{C}_1$.

The rest of the argument now follows easily from bounds on the maximum of a collection of arbitrary Gaussian random variables. Specifically, it can be seen from the previous discussion and a real-valued version of [17, Lemma 6] that

$$\Pr\left(\|(\mathbf{X}_\mathcal{S}^T \mathbf{X}_\mathcal{S})^{-1} \mathbf{X}_\mathcal{S}^T \mathbf{w}\|_\infty \geq \sqrt{2} t \big| \mathcal{C}_1\right) \leq \frac{2M e^{-t^2/2}}{\sqrt{2\pi} t}.$$

We substitute $t = \lambda/\sqrt{2}$ in the above expression to obtain

$$\frac{2M e^{-\lambda^2/4}}{\sqrt{\pi} \lambda} = \frac{1}{M(\tau+1)\sqrt{2\pi \log(M\sqrt{\tau+1})}}.$$

Summarizing, we have that the noise stability condition satisfies

$$\Pr(\mathcal{C}_2^c | \mathcal{C}_1) \leq \frac{1}{M(\tau+1)\sqrt{2\pi \log(M\sqrt{\tau+1})}}. \tag{39}$$

### C. Complementary Noise Stability

In order to establish the complementary noise-stability condition, we use ideas similar to the ones used in the previous section. To begin with, we again condition on the event $\mathcal{C}_1$ and use $\mathbf{P}_{\mathbf{X}_\mathcal{S}} \stackrel{def}{=} \mathbf{X}_\mathcal{S}(\mathbf{X}_\mathcal{S}^T \mathbf{X}_\mathcal{S})^{-1} \mathbf{X}_\mathcal{S}^T$ to denote the orthogonal projector onto the column span of $\mathbf{X}_\mathcal{S}$. Next, we use $\mathbf{z}_j$ to denote the $j$-th column of $(\mathbf{I} - \mathbf{P}_{\mathbf{X}_\mathcal{S}}) \mathbf{X}_{\mathcal{S}^c}$ and note that

$$\|\mathbf{X}_{\mathcal{S}^c}^T (\mathbf{I} - \mathbf{P}_{\mathbf{X}_\mathcal{S}}) \mathbf{w}\|_\infty = \max_{1 \leq j \leq |\mathcal{S}^c|} |\langle \mathbf{z}_j, \mathbf{w} \rangle|. \tag{40}$$

Finally, given that $\mathbf{P}_{\mathbf{X}_\mathcal{S}}$ is a projection matrix and the columns of $\mathbf{X}$ have unit norm, we have that

$$\|\mathbf{z}_j\|_2 = \|(\mathbf{I} - \mathbf{P}_{\mathbf{X}_\mathcal{S}})\mathbf{X}_{\mathcal{S}^c}\mathbf{e}_j\|_2 \leq 1, \tag{41}$$

where $\mathbf{e}_j$ denotes the $j$-th canonical basis vector.

It is now easy to see that, since $\langle \mathbf{z}_j, \mathbf{w}\rangle$ is also distributed as $\mathcal{N}(0, \|\mathbf{z}_j\|_2^2)$, we can make use of [17, Lemma 6] to obtain

$$\Pr\left(\|\mathbf{X}_{\mathcal{S}^c}^T(\mathbf{I} - \mathbf{P}_{\mathbf{X}_\mathcal{S}})\mathbf{w}\|_\infty \geq t \big| \mathcal{C}_1\right) \leq \frac{2M(\tau+1)e^{-t^2/2}}{\sqrt{2\pi} t}.$$

We substitute $t = \lambda/\sqrt{2}$ in the above expression to obtain $\frac{2M(\tau+1)e^{-\lambda^2/4}}{\sqrt{\pi}\lambda} \leq \frac{1}{M\sqrt{2\pi \log(M\sqrt{\tau+1})}}$.

Summarizing, we have that the complementary noise stability condition satisfies

$$\Pr(\mathcal{C}_3^c | \mathcal{C}_1) \leq \frac{1}{M\sqrt{2\pi \log(M\sqrt{\tau+1})}}. \tag{42}$$

*D. Size Condition*

In order to establish the size condition, we first write

$$\|(\mathbf{X}_\mathcal{S}^T\mathbf{X}_\mathcal{S})^{-1}\mathrm{sgn}(\boldsymbol{\beta}_\mathcal{S})\|_\infty \overset{(a)}{\leq} \|((\mathbf{X}_\mathcal{S}^T\mathbf{X}_\mathcal{S})^{-1} - \mathbf{I})\mathrm{sgn}(\boldsymbol{\beta}_\mathcal{S})\|_\infty + \|\mathrm{sgn}(\boldsymbol{\beta}_\mathcal{S})\|_\infty$$

$$= \|((\mathbf{X}_\mathcal{S}^T\mathbf{X}_\mathcal{S})^{-1} - \mathbf{I})\mathrm{sgn}(\boldsymbol{\beta}_\mathcal{S})\|_\infty + 1 \tag{43}$$

$$= \max_{1 \leq j \leq |\mathcal{S}|} |\langle \mathbf{z}_j, \mathrm{sgn}(\boldsymbol{\beta}_\mathcal{S})\rangle| + 1, \tag{44}$$

where $(a)$ follows from the triangle inequality and we once again use $\mathbf{z}_j$ to denote the $j$-th column of $((\mathbf{X}_\mathcal{S}^T\mathbf{X}_\mathcal{S})^{-1} - \mathbf{I})$. Now define $\mathbf{A} = (\mathbf{X}_\mathcal{S}^T\mathbf{X}_\mathcal{S} - \mathbf{I})$ and condition on the event $\mathcal{C}_1$. Then it follows from the Neumann series (cf. [6, p. 2171]) that $\|\mathbf{z}_j\|_2 \leq 2\|\mathbf{A}\mathbf{e}_j\|_2$. Furthermore, since $\mathbf{X}_\mathcal{S}$ is a submatrix of $\mathbf{X}_\mathcal{B}$, we have $\|\mathbf{A}\mathbf{e}_j\|_2 \leq \|(\mathbf{X}_\mathcal{B}^T\mathbf{X}_\mathcal{B} - \mathbf{I})\mathbf{e}_{j'}\|_2$, where $j'$ is such that the $j'$-th column of $\mathbf{X}_\mathcal{B}$ matches the $j$-th column of $\mathbf{X}_\mathcal{S}$.

Finally, define the diagonal matrix $\mathbf{Q} \overset{def}{=} \mathrm{diag}(\delta_1, \ldots, \delta_M)$ with the "random activation variables" $\{\delta_i\}$ on the diagonal and define a new matrix $\mathbf{R} = \mathbf{Q} \otimes \mathbf{I}_{\tau+1}$, where $\otimes$ denotes the Kronecker product. Next, use the notation $\mathbf{H} \overset{def}{=} (\mathbf{X}^T\mathbf{X} - \mathbf{I})$ and notice that $\|(\mathbf{X}_\mathcal{B}^T\mathbf{X}_\mathcal{B} - \mathbf{I})\mathbf{e}_{j'}\|_2 = \|\mathbf{R}\mathbf{H}\mathbf{e}_{j''}\|_2$, where $j''$ is such that the $j''$-th column of $\mathbf{X}$ matches the $j$-th column of $\mathbf{X}_\mathcal{S}$. In addition, note that $\mathbf{H} = \begin{bmatrix} \mathbf{H}_1 & \mathbf{H}_2 & \ldots & \mathbf{H}_M \end{bmatrix}$ has a block structure that can be expressed as

$$\mathbf{H} = \begin{bmatrix} \mathbf{H}_{1,1} & \mathbf{H}_{1,2} & \ldots & \mathbf{H}_{1,M} \\ \mathbf{H}_{2,1} & \mathbf{H}_{2,2} & \ldots & \mathbf{H}_{2,M} \\ \vdots & \vdots & \ddots & \vdots \\ \mathbf{H}_{M,1} & \mathbf{H}_{M,2} & \ldots & \mathbf{H}_{M,M} \end{bmatrix}, \tag{45}$$

where $\mathbf{H}_{i,j} = \mathbf{X}_i^T \mathbf{X}_j - 1_{\{i=j\}}\mathbf{I}$, $1 \leq i,j \leq M$, and $\mathbf{H}_i = [\mathbf{H}_{1,i}^T \ldots \mathbf{H}_{M,i}^T]^T$. We now define two blockwise norms on $\mathbf{H}$ as follows: $\|\mathbf{H}\|_{B,1} \stackrel{def}{=} \max_{1 \leq i \leq M} \|\mathbf{H}_i\|_2$, and $\|\mathbf{H}\|_{B,2} \stackrel{def}{=} \max_{1 \leq i,j \leq M} \|\mathbf{H}_{i,j}\|_2$.

Then it follows from the preceding discussion and the structure of the block matrix $\mathbf{H}$ that

$$\|\mathbf{z}_j\|_2 \leq 2\|\mathbf{A}\mathbf{e}_j\|_2 \leq 2\|\mathbf{R}\mathbf{H}\mathbf{e}_{j''}\|_2 \leq 2\|\mathbf{R}\mathbf{H}\|_{B,1}. \tag{46}$$

Our next goal then is to provide a bound on $\|\mathbf{R}\mathbf{H}\|_{B,1}$ and for this we resort to [18, Lemma 5].

*Proposition 2 ([18]):* For $q \geq 2\log M$ and $\delta = k/M$, we have that

$$\mathbb{E}_q\|\mathbf{R}\mathbf{H}\|_{B,1} \leq 2^{1.5}\sqrt{q}\|\mathbf{H}\|_{B,2} + \sqrt{\delta}\|\mathbf{H}\|_{B,1}. \tag{47}$$

Now notice from the definition of $\mathbf{H}$ and $\|\cdot\|_{B,2}$ that $\|\mathbf{H}\|_{B,2} \equiv \mu_B(\mathbf{X}) \leq (\tau+1)\mu(\mathbf{X})$. In addition, we have from the definition of $\mathbf{H}$ and $\|\cdot\|_{B,1}$ that

$$\|\mathbf{H}\|_{B,1} \stackrel{(b)}{\leq} \max_{1 \leq i \leq M} \|\mathbf{X}^T \mathbf{X}_i\|_2 + \|\mathbf{I}_{\tau+1}\|_2 \stackrel{(c)}{\leq} \sqrt{1+\tau\mu(\mathbf{X})}\|\mathbf{X}\|_2 + 1 \leq 2\sqrt{1+\tau\mu(\mathbf{X})}\|\mathbf{X}\|_2, \tag{48}$$

where $(b)$ follows from the definition of the spectral norm and the triangle inequality, while $(c)$ mainly follows from the fact that $\|\mathbf{X}_i\|_2 \leq \sqrt{1+\tau\mu(\mathbf{X})}$ because of the Geršgorin disc theorem [19]. We can now fix $q = 2\log M$ and make use of the above bounds to conclude from Proposition 2 that

$$\mathbb{E}_q\|\mathbf{R}\mathbf{H}\|_{B,1} \leq 4(\tau+1)\mu(\mathbf{X})\sqrt{\log M} + 2\sqrt{\delta(1+\tau\mu(\mathbf{X}))}\|\mathbf{X}\|_2. \tag{49}$$

We can now substitute (33) and (34) into the above expression to obtain $\mathbb{E}_q\|\mathbf{R}\mathbf{H}\|_{B,1} \leq \gamma_0$ with

$$\gamma_0 \stackrel{def}{=} \frac{4}{c\sqrt{\log(M(\tau+1))}} + \frac{2}{\sqrt{c\log(M(\tau+1))}}\sqrt{1+\frac{1}{c\log(M(\tau+1))}}. \tag{50}$$

In order to establish the size condition, we now define the event $\mathcal{E} = \{\max_{1 \leq j \leq |\mathcal{S}|} \|\mathbf{z}_j\|_2 < \gamma\}$ and make use of the Markov inequality along with (46) and the preceding discussion to obtain

$$\Pr(\mathcal{E}^c) \leq \gamma^{-q}\big[\mathbb{E}_q \max_{1 \leq j \leq |\mathcal{S}|} \|\mathbf{z}_j\|_2\big]^q \leq \left(\frac{2}{\gamma}\mathbb{E}_q\|\mathbf{R}\mathbf{H}\|_{B,1}\right)^q \leq \left(\frac{2\gamma_0}{\gamma}\right)^q.$$

Finally, we use $Z \stackrel{def}{=} \max_{1 \leq j \leq |\mathcal{S}|} |\langle \mathbf{z}_j, \text{sgn}(\boldsymbol{\beta}_\mathcal{S})\rangle|$ and conclude that

$$\Pr(Z \geq t) \leq \Pr(Z \geq t | \mathcal{E}) + \Pr(\mathcal{E}^c) \stackrel{(d)}{\leq} 2Me^{-t^2/2\gamma^2} + (2\gamma_0/\gamma)^q, \tag{51}$$

where $(d)$ is a consequence of the Hoeffding inequality and the union bound. The condition is now established from (43) by setting $t = 2$ in the above expression. Furthermore, set

$$\gamma = \sqrt{\frac{2}{(1+2\log 2)\log M}}, \tag{52}$$

which leads to $2Me^{-2/\gamma^2} \leq 2M^{-2\log 2}$ and

$$\frac{\gamma_0}{\gamma} \leq \frac{2(\sqrt{1+c}+2)}{0.9155c} < 1/4. \tag{53}$$

Therefore, we obtain that $\Pr(\mathcal{E}^c) \leq (1/2)^q \leq M^{-2\log 2}$ and thus we have that the size condition does not hold with probability at most

$$\Pr(\mathcal{C}_4^c|\mathcal{C}_1) \leq 3M^{-2\log 2}. \tag{54}$$

### E. Complementary Size Condition

In order to establish the complementary size condition, we proceed similar to the case of the "size condition" and define $\mathbf{z}_j$ as $\mathbf{z}_j \stackrel{def}{=} (\mathbf{X}_\mathcal{S}^T\mathbf{X}_\mathcal{S})^{-1}\mathbf{X}_\mathcal{S}^T\mathbf{X}_{\mathcal{S}^c}\mathbf{e}_j$. It can then be easily seen that $\|\mathbf{X}_{\mathcal{S}^c}^T\mathbf{X}_\mathcal{S}(\mathbf{X}_\mathcal{S}^T\mathbf{X}_\mathcal{S})^{-1}\mathrm{sgn}(\boldsymbol{\beta}_\mathcal{S})\|_\infty = \max_{1\leq j \leq |\mathcal{S}^c|} |\langle \mathbf{z}_j, \mathrm{sgn}(\boldsymbol{\beta}_\mathcal{S})\rangle|$. Now condition on the event $\mathcal{C}_1$ and notice that $\|\mathbf{z}_j\|_2 \leq 2\|\mathbf{X}_\mathcal{S}^T\mathbf{X}_{\mathcal{S}^c}\mathbf{e}_j\|_2$, $j = 1, \dots, |\mathcal{S}^c|$.

We now define $\mathbf{X}_{\mathcal{B}^c} \stackrel{def}{=} [\mathbf{X}_i : i \in \mathcal{I}^c]$ and consider the set of indices $\mathcal{T}_1 \stackrel{def}{=} \{j' : \mathbf{X}_{\mathcal{S}^c}\mathbf{e}_{j'} \text{ is a column in } \mathbf{X}_{\mathcal{B}^c}\}$. It is then easy to argue by making use of the notation developed in Section A-D that if $j \in \mathcal{T}_1$ then

$$\|\mathbf{X}_\mathcal{S}^T\mathbf{X}_{\mathcal{S}^c}\mathbf{e}_j\|_2 \leq \max_{i \in \mathcal{I}^c} \|\mathbf{X}_\mathcal{B}^T\mathbf{X}_i\|_2 = \|\mathbf{X}_\mathcal{B}^T\mathbf{X}_{\mathcal{B}^c}\|_{B,1} \stackrel{(a)}{\leq} \|\mathbf{RH}\|_{B,1}, \tag{55}$$

where $(a)$ follows from the fact that $\mathbf{X}_\mathcal{B}^T\mathbf{X}_{\mathcal{B}^c}$ is a submatrix of $\mathbf{RH}$. We therefore have from the discussion following Proposition 2 and the Markov inequality that $\forall\, j \in \mathcal{T}_1$ and for $q = 2\log M$ and $\gamma > 0$

$$\Pr(\|\mathbf{X}_\mathcal{S}^T\mathbf{X}_{\mathcal{S}^c}\mathbf{e}_j\|_2 > \gamma) \leq \frac{[\mathbb{E}_q\|\mathbf{RH}\|_{B,1}]^q}{\gamma^q} \leq \left(\frac{\gamma_0}{\gamma}\right)^q. \tag{56}$$

Finally, the argument involving $j \in \mathcal{T}_1^c$ is a little more involved but follows along similar lines. Specifically, fix any $j \in \mathcal{T}_1^c$ and define $i' \in \mathcal{I}$ to be such that $\mathbf{X}_{\mathcal{S}^c}\mathbf{e}_j$ is a column of $\mathbf{X}_{i'}$. Next, define $\tilde{\mathbf{x}}_{\mathcal{S}\cap i'}$ to be the column of $\mathbf{X}_\mathcal{S}$ that lies within the Toeplitz block $\mathbf{X}_{i'}$ and $\widetilde{\mathbf{X}}_{\mathcal{S}\setminus i'}$ to be the submatrix constructed by removing the column $\tilde{\mathbf{x}}_{\mathcal{S}\cap i'}$ from $\mathbf{X}_\mathcal{S}$. Then, if we use the notation $\mathbf{X}_{\mathcal{B}\setminus i'} \stackrel{def}{=} [\mathbf{X}_i : i \in \mathcal{B} \setminus \{i'\}]$, it can be verified that for any $j \in \mathcal{T}_1^c$ we have

$$\|\mathbf{X}_\mathcal{S}^T\mathbf{X}_{\mathcal{S}^c}\mathbf{e}_j\|_2^2 = \|\widetilde{\mathbf{X}}_{\mathcal{S}\setminus i'}^T\mathbf{X}_{\mathcal{S}^c}\mathbf{e}_j\|_2^2 + |\tilde{\mathbf{x}}_{\mathcal{S}\cap i'}^T\mathbf{X}_{\mathcal{S}^c}\mathbf{e}_j|^2$$

$$\leq \max_{i' \in \mathcal{I}} \|\mathbf{X}_{\mathcal{B}\setminus i'}^T\mathbf{X}_{i'}\|_2^2 + \mu^2(\mathbf{X})$$

$$\stackrel{(b)}{\leq} \|\mathbf{RH}\|_{B,1}^2 + \mu^2(\mathbf{X}), \tag{57}$$

where $(b)$ again makes use of the fact that the spectral norm of a matrix is an upper bound for the spectral norm of any of its submatrices. We therefore once again obtain from the discussion following

Proposition 2 and the Markov inequality that $\forall j \in \mathcal{T}_1^c$ and for $q = 2 \log M$ and $\gamma > 0$

$$\Pr(\|\mathbf{X}_\mathcal{S}^T \mathbf{X}_{\mathcal{S}^c} \mathbf{e}_j\|_2 > \gamma) \leq \Pr\left(\|\mathbf{R}\mathbf{H}\|_{B,1} > \sqrt{\gamma^2 - \mu^2(\mathbf{X})}\right) \leq \left(\frac{\gamma_0}{\sqrt{\gamma^2 - \mu^2(\mathbf{X})}}\right)^q. \quad (58)$$

We can now define the event $\mathcal{E} = \{\|\mathbf{X}_\mathcal{S}^T \mathbf{X}_{\mathcal{S}^c} \mathbf{e}_j\|_2 \leq \gamma\}$ and use the notation $Z \stackrel{def}{=} \max_{1 \leq j \leq |\mathcal{S}^c|} |\langle \mathbf{z}_j, \text{sgn}(\boldsymbol{\beta}_\mathcal{S})\rangle|$ to conclude from (56) and (58) that

$$\Pr(Z \geq t) \leq \Pr(Z \geq t|\mathcal{E}) + \Pr(\mathcal{E}^c)$$
$$\stackrel{(c)}{\leq} 2M(\tau+1)e^{-t^2/2\gamma^2} + (\gamma_0/\gamma)^q + (\gamma_0/\sqrt{\gamma^2 - \mu^2(\mathbf{X})})^q, \quad (59)$$

where $(c)$ follows from the Hoeffding inequality and the union bound. The condition is now established by setting $t = \frac{1}{4}$ in the above expression. Furthermore, set

$$\gamma = \frac{1}{\sqrt{32(1 + 2\log 2)\log(M(\tau+1))}}, \quad (60)$$

which yields $2M(\tau+1)e^{-1/32\gamma^2} \leq 2(M(\tau+1))^{-2\log 2}$ and $\frac{\gamma_0}{\sqrt{\gamma^2-\mu^2}} \leq \frac{\frac{2\sqrt{1+c}}{c} + \frac{4}{c}}{\sqrt{0.1144^2 - 1/c^2}} < 1/2$. Therefore, we obtain that $\Pr(\mathcal{E}^c) \leq 2(\gamma_0/\sqrt{\gamma^2-\mu^2})^q \leq 2(1/2)^q \leq 2(M(\tau+1))^{-2\log 2}$ and thus we have that the size condition satisfies

$$\Pr(\mathcal{C}_5^c|\mathcal{C}_1) \leq 4\big(M(\tau+1)\big)^{-2\log 2}. \quad (61)$$

*F. Proof of Theorem 1*

The proof of Theorem 1 follows from the preceding discussion by taking a union bound over all the respective conditions and removing the conditionings: $\Pr((\mathcal{C}_1 \cap \mathcal{C}_2 \cap \mathcal{C}_3 \cap \mathcal{C}_4 \cap \mathcal{C}_5)^c) \leq \Pr(\mathcal{C}_1^c) + \Pr(\mathcal{C}_2^c|\mathcal{C}_1) + \Pr(\mathcal{C}_3^c|\mathcal{C}_1) + \Pr(\mathcal{C}_4^c|\mathcal{C}_1) + \Pr(\mathcal{C}_5^c|\mathcal{C}_1)$. Consequently, we obtain that the probability of error is upper bounded by $2M^{-1}\big(2\pi \log(M\sqrt{\tau+1})\big)^{-1/2} + 5\big(M(\tau+1)\big)^{-2\log 2} + 3M^{-2\log 2}$.

# APPENDIX B
# PROOF OF THE MAIN RESULT: RANDOM DELAYS

In this appendix, we provide a proof of Theorem 2. The proof parallels that of Theorem 1, thus the definitions and notation in Appendix A are reused. Key to the proof is the distribution and generation of the support set $\mathcal{S}$, which we examine first.

As described in Section 2, here we consider the case when $\tau_i$ are uniformly selected from $\{0, \ldots, \tau\}$ at random. Translating the notions of users and delays to the block structure of $\mathbf{X}$, the set $\mathcal{S}$ can be viewed as generated by a two step procedure: (1) blocks are activated with probability $\delta = k/M$; (2) within each active block, a delay/column is selected uniformly at random. We call this the *conventional* *activation procedure* (CAP). However, to prove Theorem 2, it is useful to examine a different activation procedure of $\mathcal{S}$ as follows:

1) Let $\{\tilde{\delta}_i\}_{i=1}^{M(\tau+1)}$ be a set of Bernoulli random variables with $\Pr(\tilde{\delta}_i = 1) = \frac{k\rho}{M(\tau+1)} \stackrel{def}{=} \tilde{\delta}$. Set $\tilde{\mathcal{S}}$ to be the set $\{i : \tilde{\delta}_i = 1\}$.
2) Mapping indices to the block structure on $\mathbf{X}$, prune $\tilde{\mathcal{S}}$ to $\mathcal{S}$: For each block with more than one active element in $\tilde{\mathcal{S}}$, select a single element uniformly at random among the active elements in the block.

We call this the *equivalent activation procedure* (EAP) and we now argue that, with an appropriate value of $\rho$, the set $\mathcal{S}$ is distributed identically to that generated using the conventional procedure. Of particular utility will be the set $\tilde{\mathcal{S}}$ since $\tilde{\mathcal{S}} \supset \mathcal{S}$ and $\tilde{\mathcal{S}}$ is generated simply from iid Bernoulli variables. We further define $\tilde{\mathcal{S}}_i$ to be $\tilde{\mathcal{S}}$ restricted to elements in block $i$.

The value of $\rho$ needed can be calculated by requiring the probability of block activity to be equal under the CAP and the EAP. That is, for any $i = 1, \ldots, M$,

$$\Pr[\tilde{\mathcal{S}}_i > 1] = 1 - \Pr[\tilde{\mathcal{S}}_i = 0] = 1 - \left(1 - \frac{k\rho}{M(\tau+1)}\right)^{\tau+1} = \frac{k}{M}, \quad (62)$$

where the last equality links the two procedures. Solving for $\rho$ gives

$$\rho = \left(1 - \left(1 - \frac{k}{M}\right)^{\frac{1}{\tau+1}}\right) \frac{M(\tau+1)}{k}. \quad (63)$$

When $k \ll M$, $\left(1 - \frac{k}{M}\right)^{1/(\tau+1)} \approx 1 - \frac{k}{M(\tau+1)}$ and $\rho \approx 1$ as expected. This approximation will be made more explicit later in (67).

To prove equivalence in distribution between the two methods, it remains to show independence of blocks and uniformity among columns in blocks in the EAP. Independence of blocks is inherited from the independence of column activation in Step 1 (since the blocks are disjoint sets). We now make a symmetry argument to show a uniform selection of columns.

Let $(i, j)$ be an arbitrary column/block pair. Let $\mathcal{Y}$ be the event that $(i, j)$ is activated in Step 1 and let $\mathcal{X}$ be the event that $(i, j)$ is selected in Step 2. Since the events satisfy $\mathcal{X} \subset \mathcal{Y} \subset \{|\tilde{\mathcal{S}}_j| > 0\}$, we can

write

$$\begin{aligned}
\Pr[\mathcal{X}] &= \Pr[\mathcal{X} \cap \mathcal{Y} \cap \{|\tilde{\mathcal{S}}_j| > 0\}] \\
&= \sum_{n=1}^{\tau+1} \Pr[\mathcal{X} \cap \mathcal{Y} \cap \{|\tilde{\mathcal{S}}_j| = n\}] \\
&= \sum_{n=1}^{\tau+1} \Pr[\mathcal{X} \big| \mathcal{Y} \cap \{|\tilde{\mathcal{S}}_j| = n\}] \Pr[\mathcal{Y} \cap \{|\tilde{\mathcal{S}}_j| = n\}] \\
&= \sum_{n=1}^{\tau+1} \frac{1}{n} \Pr[\mathcal{Y} \cap \{|\tilde{\mathcal{S}}_j| = n\}],
\end{aligned} \qquad (64)$$

where the last equality is due to the uniform selection in Step 2. Now, for $n = 1, \ldots, \tau+1$, we have

$$\begin{aligned}
\Pr[\mathcal{Y} \cap \{|\tilde{\mathcal{S}}_j| = n\}] &= \Pr\left[\{|\tilde{\mathcal{S}}_j| = n\} \big| \mathcal{Y}\right] \Pr[\mathcal{Y}] \\
&= \left[\binom{\tau}{n-1} \left(1 - \frac{k\rho}{(\tau+1)M}\right)^{\tau-n+1} \left(\frac{k\rho}{(\tau+1)M}\right)^{n-1}\right] \left[\frac{k\rho}{(\tau+1)M}\right] \\
&= \binom{\tau}{n-1} \left(1 - \frac{k\rho}{(\tau+1)M}\right)^{\tau-n+1} \left(\frac{k\rho}{(\tau+1)M}\right)^{n},
\end{aligned} \qquad (65)$$

where the first factor is Binomial over the $\tau$ remaining columns given that $i, j$ was selected in Step 1. At this point, it is sufficient to note that (65) is not a function of our choice of $(i, j)$. Thus, by symmetry, the probability is equal for all columns. Nonetheless, we complete the calculation to show it takes the anticipated value. Returning to (64), we have,

$$\begin{aligned}
\Pr[\mathcal{X}] &= \sum_{n=1}^{\tau+1} \frac{1}{n} \binom{\tau}{n-1} \left(1 - \frac{k\rho}{(\tau+1)M}\right)^{\tau-n+1} \left(\frac{k\rho}{(\tau+1)M}\right)^{n} \\
&= \sum_{n=1}^{\tau+1} \frac{1}{\tau+1} \binom{\tau+1}{n} \left(1 - \frac{k\rho}{(\tau+1)M}\right)^{\tau-n+1} \left(\frac{k\rho}{(\tau+1)M}\right)^{n} \\
&= \frac{1}{\tau+1} \left[1 - \left(1 - \frac{k\rho}{(\tau+1)M}\right)^{\tau+1}\right] \\
&= \frac{1}{\tau+1} \frac{k}{M},
\end{aligned}$$

where in the second equality we use a simple identity on Binomial coefficient, in the third equality we note the sum is nearly complete over a Binomial distribution function and lastly, we use (62).

Having shown the equivalence between CAP and EAP, we are ready to prove the five conditions $\mathcal{C}_1$ through $\mathcal{C}_5$ that guarantee recovery. While our model on the users corresponds to the CAP, we will use the EAP in the remainder of the proof. Since $\tilde{\mathcal{S}}$ is formed from iid random variables, we are able to follow a proof technique similar to that of [6]. We include our proof for completeness since our theory is based on slightly different assumptions and aspects of our proof use different methods.

## A. Invertibility Condition

To bound $\|(\mathbf{X}_\mathcal{S}^T\mathbf{X}_\mathcal{S})^{-1}\|_2$ we consider $\mathcal{S}$ and $\tilde{\mathcal{S}}$ as generated from EAP. Since $\tilde{\mathcal{S}}$ is uniformly distributed over possible column selections, we can bound $\|(\mathbf{X}_{\tilde{\mathcal{S}}}^T\mathbf{X}_{\tilde{\mathcal{S}}})^{-1}\|_2$ using methods of [6] where, using [20] and $q = 2\log(M(\tau+1))$, we have

$$\mathbb{E}_q\|\mathbf{X}_{\tilde{\mathcal{S}}}^T\mathbf{X}_{\tilde{\mathcal{S}}} - \mathbf{I}\|_2 \leq 30\mu(\mathbf{X})\log(M(\tau+1)) + 13\sqrt{\frac{2k\rho\|\mathbf{X}\|_2^2\log(M(\tau+1))}{M(\tau+1)}}. \tag{66}$$

We would like to translate this into conditions similar to (33) and (34). To do so, we make the approximation noted below (63) explicit. We will assume $k/M \leq 1/4$ here, which follows trivially from the condition (14) in the theorem. This assumption allows us to make the following approximation.

$$1 - (1 - k/M)^{1/(\tau+1)} \leq \frac{k}{M(\tau+1)}(1 - 1/4)^{\frac{1}{\tau+1}-1} \leq \frac{k}{M(\tau+1)}\frac{4}{3} \tag{67}$$

The first inequality is an application of Taylor's remainder theorem on the function $f(x) = 1 - (1-x)^{1/(\tau+1)}$, while the second inequality is due to the fact that $(1-\epsilon)^{\frac{1}{\tau+1}} \leq 1$ for $\epsilon \geq 0$. Applying this approximation to (63) yields $\rho \leq 4/3$.

Thus, if

$$\mu(\mathbf{X}) \leq \frac{1}{c'\log(M(\tau+1))} \quad \text{and} \tag{68}$$

$$\|\mathbf{X}\|_2^2 \leq \frac{M(\tau+1)}{c'k\log(M(\tau+1))} \leq \frac{M(\tau+1)}{c''k\rho\log(M(\tau+1))}, \tag{69}$$

then $\mathbb{E}_q\|\mathbf{X}_{\tilde{\mathcal{S}}}^T\mathbf{X}_{\tilde{\mathcal{S}}} - \mathbf{I}\|_2 \leq 1/4$. Above, $c'$ and $c''$ are appropriately chosen constants independent of the problem parameters. Since $\mathcal{S} \subset \tilde{\mathcal{S}}$, we have $\|\mathbf{X}_\mathcal{S}^T\mathbf{X}_\mathcal{S} - \mathbf{I}\|_2 \leq \|\mathbf{X}_{\tilde{\mathcal{S}}}^T\mathbf{X}_{\tilde{\mathcal{S}}} - \mathbf{I}\|_2$ and therefore $\mathbb{E}_q\|\mathbf{X}_\mathcal{S}^T\mathbf{X}_\mathcal{S} - \mathbf{I}\|_2 \leq 1/4$. Following the calculations in Appendix A-A, cf. (36)-(37), this gives

$$\Pr(\mathcal{C}_1^c) \leq (M(\tau+1))^{-2\log 2}. \tag{70}$$

## B. Noise Stability and Complementary Noise Stability

Conditions $\mathcal{C}_2$ and $\mathcal{C}_3$ follow when conditioned on $\mathcal{C}_1$ in an identical manner to Appendix A-B and Appendix A-C with probabilities (39) and (42), respectively.

## C. Size Condition

For $\mathcal{C}_4$ we begin as in Appendix A-D with the following upper bound

$$\|(\mathbf{X}_\mathcal{S}^T\mathbf{X}_\mathcal{S})^{-1}\mathrm{sgn}(\boldsymbol{\beta}_\mathcal{S})\|_\infty \leq \max_{1\leq j \leq |\mathcal{S}|} |\langle \mathbf{z}_j, \mathrm{sgn}(\boldsymbol{\beta}_\mathcal{S})\rangle| + 1,$$

where $\mathbf{z}_j$ denotes the $j$-th column of $((\mathbf{X}_\mathcal{S}^T\mathbf{X}_\mathcal{S})^{-1} - \mathbf{I})$. Using the definitions from Section A-D and additionally defining $\tilde{\mathbf{A}} = \mathbf{X}_{\tilde{\mathcal{S}}}^T \mathbf{X}_{\tilde{\mathcal{S}}} - \mathbf{I}$, and conditioning on $\mathcal{C}_1$, we have

$$\|\mathbf{z}_j\|_2 \leq 2\|\mathbf{A}\mathbf{e}_j\|_2 \leq 2\|\tilde{\mathbf{A}}\mathbf{e}_{j'}\|_2, \tag{71}$$

where the first inequality is due to an application of the Neumann series and $\mathcal{C}_1$. The second inequality is due to $\mathbf{A}$ being a sub-matrix of $\tilde{\mathbf{A}}$ and the choice of $j'$ such that the column $\mathbf{X}_\mathcal{S}\mathbf{e}_j$ is the same as $\mathbf{X}_{\tilde{\mathcal{S}}}\mathbf{e}_{j'}$.

Next we define $\tilde{\mathbf{R}} \stackrel{def}{=} \mathrm{diag}(\tilde{\delta}_1,\ldots,\tilde{\delta}_{M(\tau+1)})$ as a selection matrix for the EAP similar to $\mathbf{R}$ so that $\mathbf{X}_{\tilde{\mathcal{S}}} = \tilde{\mathbf{R}}\mathbf{X}$ (conforming to the first step in the EAP). With this definition we have

$$\|\tilde{\mathbf{A}}\mathbf{e}_{j'}\|_2 \leq \|\tilde{\mathbf{R}}\mathbf{H}\|_{1\to 2}, \tag{72}$$

where $\|\cdot\|_{1\to 2}$ denotes the maximal column norm as defined in [20].

Since $\{\tilde{\delta}_i\}$ are iid Bernoulli random variables, we can apply [20, Theorem 3.2] which gives, for $q = 2\log(M(\tau+1))$,

$$\mathbb{E}_q\|\tilde{\mathbf{R}}\mathbf{H}\|_{1\to 2} \leq 2^{1.75}\sqrt{\log(M(\tau+1))}\mu(\mathbf{X}) + \sqrt{\tilde{\delta}}\|\mathbf{H}\|_{1\to 2}$$

We can bound $\|\mathbf{H}\|_{1\to 2}$ as follows:

$$\|\mathbf{H}\|_{1\to 2} = \max_{1\leq i \leq M(\tau+1)} \|(\mathbf{X}^T\mathbf{X} - \mathbf{I})\mathbf{e}_i\|_2 \leq \max_{1\leq i \leq M(\tau+1)} \|(\mathbf{X}^T\tilde{\mathbf{x}}_i)\|_2 + 1$$

$$\leq \|\mathbf{X}\|_2 + 1 \leq 2\|\mathbf{X}\|_2$$

where we use $1 \leq \|\mathbf{X}\|_2$ since the columns have unit norm. This gives

$$\mathbb{E}_q\|\tilde{\mathbf{R}}\mathbf{H}\|_{1\to 2} \leq 2^{1.75}\sqrt{\log(M(\tau+1))}\mu(\mathbf{X}) + 2\sqrt{\tilde{\delta}}\|\mathbf{X}\|_2.$$

Upon substituting in the values from (33) and (34), we obtain

$$\mathbb{E}_q\|\tilde{\mathbf{R}}\mathbf{H}\|_{1\to 2} \leq \tilde{\gamma}_0 \tag{73}$$

with

$$\tilde{\gamma}_0 = \frac{1}{\sqrt{\log(M(\tau+1))}}\left[\frac{2^{1.75}}{c} + \frac{2}{\sqrt{c}}\right] \tag{74}$$

In order to establish the size condition, we now define the event $\mathcal{E} = \{\max_{1\leq j\leq |\mathcal{S}|} \|\mathbf{z}_j\|_2 < \gamma\}$ and make use of the Markov inequality along with (71), (72) and the preceding discussion to obtain

$$\Pr(\mathcal{E}^c) \leq \gamma^{-q} \left[\mathbb{E}_q \max_{1\leq j\leq |\mathcal{S}|} \|\mathbf{z}_j\|_2\right]^q \leq \left(\frac{2}{\gamma}\mathbb{E}_q\|\tilde{\mathbf{R}}\mathbf{H}\|_{1\to 2}\right)^q \leq \left(\frac{2\tilde{\gamma}_0}{\gamma}\right)^q. \tag{75}$$

Finally, we use $Z \stackrel{def}{=} \max_{1\leq j\leq |\mathcal{S}|} |\langle \mathbf{z}_j, \mathrm{sgn}(\boldsymbol{\beta}_{\mathcal{S}})\rangle|$ and conclude that

$$\Pr(Z \geq t) \leq \Pr(Z \geq t|\mathcal{E}) + \Pr(\mathcal{E}^c) \stackrel{(a)}{\leq} 2M(\tau+1)e^{-t^2/2\gamma^2} + (2\tilde{\gamma}_0/\gamma)^q, \tag{76}$$

where $(a)$ is a consequence of the Hoeffding inequality and the union bound. The condition is now established from (43) by setting $t = 2$ in the above expression. Furthermore, set

$$\gamma = \sqrt{\frac{2}{(1+2\log 2)\log(M(\tau+1))}}, \tag{77}$$

which leads to $2M(\tau+1)e^{-2/\gamma^2} \leq 2\left(M(\tau+1)\right)^{-2\log 2}$ and

$$\frac{\tilde{\gamma}_0}{\gamma} \leq \frac{(2^{1.75} + 2\sqrt{c})(1+2\log 2)}{c} < 1/4. \tag{78}$$

Therefore, we obtain that $\Pr(\mathcal{E}^c) \leq (1/2)^q \leq (M(\tau+1))^{-2\log 2}$ and thus we have that the size condition does not hold with probability at most

$$\Pr(\mathcal{C}_4^c|\mathcal{C}_1) \leq 3\left(M(\tau+1)\right)^{-2\log 2}.$$

*D. Complementary Size Condition*

As in Appendix A-E, we define $\mathbf{z}_j$ as $\mathbf{z}_j \stackrel{def}{=} (\mathbf{X}_{\mathcal{S}}^T\mathbf{X}_{\mathcal{S}})^{-1}\mathbf{X}_{\mathcal{S}}^T\mathbf{X}_{\mathcal{S}^c}\mathbf{e}_j$. It can then be easily seen that $\|\mathbf{X}_{\mathcal{S}^c}^T\mathbf{X}_{\mathcal{S}}(\mathbf{X}_{\mathcal{S}}^T\mathbf{X}_{\mathcal{S}})^{-1}\mathrm{sgn}(\boldsymbol{\beta}_{\mathcal{S}})\|_{\infty} = \max_{1\leq j\leq |\mathcal{S}^c|} |\langle \mathbf{z}_j, \mathrm{sgn}(\boldsymbol{\beta}_{\mathcal{S}})\rangle|$. Now condition on the event $\mathcal{C}_1$ and notice that $\|\mathbf{z}_j\|_2 \leq 2\|\mathbf{X}_{\mathcal{S}}^T\mathbf{X}_{\mathcal{S}^c}\mathbf{e}_j\|_2$, $j = 1,\ldots,|\mathcal{S}^c|$.

We then see that

$$\|\mathbf{X}_{\mathcal{S}}^T\mathbf{X}_{\mathcal{S}^c}\mathbf{e}_j\|_2 \leq \|\mathbf{X}_{\mathcal{S}}^T\mathbf{X}_{\mathcal{S}^c}\|_{1\to 2} \stackrel{(a)}{\leq} \|\tilde{\mathbf{R}}\mathbf{H}\|_{1\to 2}, \tag{79}$$

where $(a)$ follows from the fact that $\mathbf{X}_{\mathcal{S}}^T\mathbf{X}_{\mathcal{S}^c}$ is a submatrix of $\tilde{\mathbf{R}}\mathbf{H}$. As in the previous subsection, by redefining the event $\mathcal{E} = \{\max_{1\leq j\leq |\mathcal{S}^c|} \|\mathbf{z}_j\|_2 < \gamma\}$ and $Z \stackrel{def}{=} \max_{1\leq j\leq |\mathcal{S}^c|} |\langle \mathbf{z}_j, \mathrm{sgn}(\boldsymbol{\beta}_{\mathcal{S}})\rangle|$ and using (73), (75) and (76) hold once again.

In accordance with the complementary size condition, we take $t = 1/4$ and set

$$\gamma = \sqrt{\frac{1}{32(1+2\log 2)\log(M(\tau+1))}} \tag{80}$$

so that $2M(\tau+1)e^{-\frac{t^2}{2\gamma^2}} = 2(M(\tau+1))^{-2\log 2}$ and $\tilde{\gamma}_0/\gamma \leq 1/4$. This gives us

$$\Pr(\mathcal{C}_5^c|\mathcal{C}_1) \leq 3(M(\tau+1))^{-2\log 2}.$$

*E. Proof of Theorem 2*

The proof of Theorem 2 follows from the preceding discussion by taking a union bound over all the respective conditions and removing the conditionings: $\Pr((\mathcal{C}_1 \cap \mathcal{C}_2 \cap \mathcal{C}_3 \cap \mathcal{C}_4 \cap \mathcal{C}_5)^c) \leq \Pr(\mathcal{C}_1^c) + \Pr(\mathcal{C}_2^c|\mathcal{C}_1) + \Pr(\mathcal{C}_3^c|\mathcal{C}_1) + \Pr(\mathcal{C}_4^c|\mathcal{C}_1) + \Pr(\mathcal{C}_5^c|\mathcal{C}_1)$. Consequently, we obtain that the probability of error is upper bounded by $2M^{-1}\big(2\pi \log(M\sqrt{\tau+1}\,)\big)^{-1/2} + 7\big(M(\tau+1)\big)^{-2\log 2}$.

# APPENDIX C
## PROOF OF LEMMA 3

In order to bound the sum of (25), we will use the following propositions.

*Proposition 3:* For $x, y \in \mathrm{GF}(2^m)$ and $i = 1, 2, \ldots$

$$x^{2^i+1} + y^{2^i+1} = (x+y)^{2^i+1} + \sum_{j=0}^{i-1}(xy)^{2^j}(x+y)^{2^i - 2^{j+1}+1}$$

*Proof:* We prove this by induction and application of $(x+y)^{2^i} = x^{2^i} + y^{2^i}$. That is, we first note that the lemma holds for $i = 1$ and, assuming true for $i$, we have

$$(x+y)^{2^{i+1}+1} = (x+y)^{2^i}(x+y)^{2^i+1}$$

$$= (x+y)^{2^i}\Big[x^{2^i+1} + y^{2^i+1} + \sum_{j=0}^{i-1}(xy)^{2^j}(x+y)^{2^i - 2^{j+1}+1}\Big]$$

$$= (x^{2^i} + y^{2^i})(x^{2^i+1} + y^{2^i+1}) + \sum_{j=0}^{i-1}(xy)^{2^j}(x+y)^{2^{i+1} - 2^{j+1}+1}$$

$$= x^{2^{i+1}+1} + y^{2^{i+1}+1} + (xy)^{2^i}(x+y) + \sum_{j=0}^{i-1}(xy)^{2^j}(x+y)^{2^{i+1} - 2^{j+1}+1}.$$

Incorporating the middle term in the sum completes the proof. ■

*Proposition 4 ([14, pp. 278–9]):* The quadratic polynomial $x^2 + fx + g$ with coefficients in $\mathrm{GF}(2^m)$ and $f \neq 0$ has two distinct roots in $\mathrm{GF}(2^m)$ if $\mathrm{Tr}(g/f^2) = 0$ and no roots in $\mathrm{GF}(2^m)$ if $\mathrm{Tr}(g/f^2) = 1$.

*Proposition 5:* (a) The cardinality of $\{g \in \mathrm{GF}(2^m) : \mathrm{Tr}(\alpha g) = c_1\}$ is $2^{m-1}$ for $\alpha \in \mathrm{GF}(2^m), \alpha \neq 0$ and $c_1 \in \mathrm{GF}(2)$. (b) The cardinality of $\{g \in \mathrm{GF}(2^m) : \mathrm{Tr}(\alpha g) = c_1, \mathrm{Tr}(\beta g) = c_2\}$ is $2^{m-2}$ for $\beta \in \mathrm{GF}(2^m), \beta \neq \alpha, \beta \neq 0$ and $c_2 \in \mathrm{GF}(2)$.

*Proof:* Let $\{\eta_i\}_{i=1}^m$ and $\{\lambda_i\}_{i=1}^m$ be dual bases of $\mathrm{GF}(2^m)$ [14, p. 117] and consider $\alpha$ and $g$ in these bases respectively as $\alpha = a_1\eta_1 + \cdots + a_m\eta_m$ and $g = \gamma_1\lambda_1 + \cdots + \gamma_m\lambda_m$ for $a_i, \gamma_i \in \mathrm{GF}(2)$. Then $\mathrm{Tr}(\alpha g) = a_1\gamma_1 + \cdots + a_m\gamma_m = c_1$ is a restriction of a single degree of freedom in selecting $\{\gamma_i\}_{i=1}^m$. Similarly, $\mathrm{Tr}(\beta g) = c_2$ restricts an additional degree of freedom. ■

We are now ready to bound the sum given by (25). We will begin with a simple (yet required) case which illustrates our use of Proposition 5. Suppose the non-zero vector $\alpha$ is zero everywhere but at $\alpha_0$. In this case we have

$$S = \sum_{x \in \mathrm{GF}^*(2^m)} (-1)^{\mathrm{Tr}(\alpha_0 x)} = \sum_{x \in \mathrm{GF}(2^m)} (-1)^{\mathrm{Tr}(\alpha_0 x)} - 1$$

where we've completed the sum to be over all of $\mathrm{GF}(2^m)$. By Proposition 5 (a), $\mathrm{Tr}(\alpha_0 x) = 1$ for precisely presicely half the $2^m$ terms of the sum. Thus, the sum is 0 and $|S| = 1$. For the remainder of the proof, we will assume that $\alpha_i$ is non-zero for some $i \geq 1$.

Considering the square of (25), by using the linearity of the trace we have

$$S^2 = \sum_{x \in \mathrm{GF}^*(2^m)} \sum_{y \in \mathrm{GF}^*(2^m)} (-1)^{\mathrm{Tr}\left[\alpha_0(x+y) + \sum_{i=1}^{t} \alpha_i(x^{2^i+1} + y^{2^i+1})\right]}$$

$$= 2^m - 1 + \sum_{\substack{x \in \mathrm{GF}^*(2^m) \\ y \neq x}} \sum_{y \in \mathrm{GF}^*(2^m)} (-1)^{\mathrm{Tr}\left[\alpha_0(x+y) + \sum_{i=1}^{t} \alpha_i(x^{2^i+1} + y^{2^i+1})\right]}$$

$$= 2^m - 1 + \sum_{\substack{x \in \mathrm{GF}^*(2^m) \\ y \neq x}} \sum_{y \in \mathrm{GF}^*(2^m)} (-1)^{\mathrm{Tr}\left[\alpha_0(x+y) + \sum_{i=1}^{t} \alpha_i((x+y)^{2^i+1} + \sum_{j=0}^{i-1}(xy)^{2^j}(x+y)^{2^i-2^{j+1}+1})\right]}.$$

In the last equality we have used Proposition 3 so that we may apply the change of variables given by $f = x + y$ and $g = xy$. To justify this substitution we note that

$$\{(x + y, xy) : x \in \mathrm{GF}^*(2^m), y \in \mathrm{GF}^*(2^m), y \neq x\}$$
$$= \{(f, g) : f \in \mathrm{GF}^*(2^m), g \in \mathrm{GF}^*(2^m), \mathrm{Tr}(g/f^2) = 0\}.$$

To see this, consider quadratics $(z+x)(z+y) = z^2 + fz + g$ with non-zero roots. The first set generates all quadratics with two solutions by enumerating the roots while, by Proposition 4, the second set generates the same quadratics by enumerating the coefficients. Since with this substitution both $(x, y)$ and $(y, x)$ map to $(f, g)$ we account for the extra factor of 2 below. We now have

$$
\begin{aligned}
S^2 &= 2^m - 1 + 2 \sum_{\substack{f,g \in \mathrm{GF}^*(2^m) \\ \mathrm{Tr}(g/f^2)=0}} (-1)^{\mathrm{Tr}\left[\alpha_0 f + \sum_{i=1}^{t} \alpha_i (f^{2^i+1} + \sum_{j=0}^{i-1} g^{2^j} f^{2^i - 2^{j+1}+1})\right]} \\
&= 2^m - 1 + 2 \sum_{f \in \mathrm{GF}^*(2^m)} (-1)^{\mathrm{Tr}\left(\alpha_0 f + \sum_{i=1}^{t} \alpha_i f^{2^i+1}\right)} \sum_{\substack{g \in \mathrm{GF}^*(2^m) \\ \mathrm{Tr}(g/f^2)=0}} (-1)^{\mathrm{Tr}\left(\sum_{i=1}^{t}\sum_{j=0}^{i-1} \alpha_i g^{2^j} f^{2^i - 2^{j+1}+1}\right)} \\
&= 2^m - 1 + 2 \sum_{f \in \mathrm{GF}^*(2^m)} (-1)^{\mathrm{Tr}\left(\alpha_0 f + \sum_{i=1}^{t} \alpha_i f^{2^i+1}\right)} \Bigg[ \sum_{\substack{g \in \mathrm{GF}(2^m) \\ \mathrm{Tr}(g/f^2)=0}} (-1)^{\mathrm{Tr}\left(\sum_{i=1}^{t}\sum_{j=0}^{i-1} \alpha_i g^{2^j} f^{2^i - 2^{j+1}+1}\right)} - 1 \Bigg] \\
&\leq 3(2^m - 1) + 2 \sum_{f \in \mathrm{GF}^*(2^m)} (-1)^{\mathrm{Tr}\left(\alpha_0 f + \sum_{i=1}^{t} \alpha_i f^{2^i+1}\right)} \sum_{\substack{g \in \mathrm{GF}(2^m) \\ \mathrm{Tr}(g/f^2)=0}} (-1)^{\mathrm{Tr}\left(\sum_{i=1}^{t}\sum_{j=0}^{i-1} \alpha_i g^{2^j} f^{2^i - 2^{j+1}+1}\right)}
\end{aligned}
$$
(81)

where, in the third equality, we've completed the sum in $g$ to include $g = 0$. The resulting subtraction creates a sum over $f$ which we trivially bound by $2^m - 1$. Turning our attention to the innermost sum over $g$, we will show that the sum is either 0 or $2^{m-1}$. Further, we will bound the number of $f$ for which it is not zero.

To separate $g$, we can use the linearity of the trace and $\mathrm{Tr}(x) = \mathrm{Tr}(x^{2^{-j}})$ for each $j$ and rewrite the exponent of the inner sum as $\mathrm{Tr}\big[(\sum_{i=1}^{t}\sum_{j=0}^{i-1} \alpha_i^{2^{-j}} f^{2^{i-j}+2^{-j}-2})g\big] = \mathrm{Tr}(\Gamma_f g)$ where we've introduced $\Gamma_f \in \mathrm{GF}(2^m)$ to simplify notation. Suppose, for a fixed $f$, there exists some $g$ with $\mathrm{Tr}(g/f^2) = 0$ such that $(-1)^{\mathrm{Tr}(\Gamma_f g)} = -1$. Then we must have $\Gamma_f \neq 1/f^2$ and $\Gamma_f \neq 0$. In this case, Proposition 5 tells us that the inner sum of (81) evaluates to 0 since part (a) gives the size of the sum while (b) shows precisely half the terms take value (-1). Thus, we are interested in when $(-1)^{\mathrm{Tr}(\Gamma_f g)}$ maps all of the subset $\{g \in \mathrm{GF}(2^m) : \mathrm{Tr}(g/f^2) = 0\}$ to 1.

When $(-1)^{\mathrm{Tr}(\Gamma_f g)}$ is a trivial map of the subset, we have $\{g : \mathrm{Tr}(\Gamma_f g) = 0\} \supseteq \{g : \mathrm{Tr}(g/f^2) = 0\}$ which provides two cases. The first is that $\Gamma_f = 0$ and the above inclusion is strict. In second case, when $\Gamma_f \neq 0$, the sets have same cardinality by Proposition 5 and, thus, the two sets are equal. In this case, by $\mathrm{Tr}\big[(\Gamma_f + 1/f^2)g\big] = 0 \quad \forall g$, the non-degeneracy of the trace [21, Proposition 28.87] tells us $\Gamma_f = 1/f^2$. In both cases, Proposition 5 gives the size of the inner sum of (81) as $2^{m-1}$. The task now becomes to bound the number of $f$ for which each of these cases occur.

$\Gamma_f = 0$ defines the following polynomial in $f$:

$$
0 = \sum_{i=1}^{t} \sum_{j=0}^{i-1} \alpha_i^{2^{-j}} f^{2^{i-j}+2^{-j}-2} = \sum_{i=1}^{t} \sum_{j=0}^{i-1} \alpha_i^{2^{t-j-1}} f^{2^{t+i-j-1}+2^{t-j-1}-2^t}
$$

where, in the second equality, we've used $(x+y)^{2^{t-1}} = x^{2^{t-1}} + y^{2^{t-1}}$ to ensure the powers of $f$ are positive integers. The degree of this polynomial is at most $2^{2t-1} + 2^{t-1} - 2^t$ and thus we at most $2^{2t-1} + 2^{t-1} - 2^t$ roots at which $\Gamma_f = 0$.

The case for $f^2 \Gamma_f = 1$ is similar and follows the same steps on a slightly different polynomial. In this case we find there are at most $2^{2t-1} + 2^{t-1}$ values of $f$ for which $\Gamma_f = 1/f^2$. Combining the two cases, we find that there are at most $2^{2t}$ values of $f$ for which $(-1)^{\mathrm{Tr}(\Gamma_f g)}$ is a trivial map over the sum. Returning to (81), we've found the sum in $f$ have terms with values of either $0$ or $\pm 2^{m-1}$ with the non-zeros terms occurring at most $2^{2t}$ times. Thus,

$$S^2 \leq 3(2^m - 1) + 2 \times 2^{2t} \times 2^{m-1} \leq 2^{m+2t+1}$$

Taking the root gives the result.

## REFERENCES

[1] S. Verdú, *Multiuser Detection*. Cambridge, U.K.: Cambridge University Press, 1998.

[2] A. Fletcher, S. Rangan, and V. Goyal, "On-off random access channels: A compressed sensing framework," submitted [arXiv:0903.1022v2].

[3] M. Buehrer, *Code Division Multiple Access (CDMA)*. Morgan & Claypool, 2006.

[4] L. Zhang, J. Luo, and D. Guo, "Compressed neighbor discovery for wireless networks," submitted. [Online]. Available: arXiv:1012.1007v2

[5] L. Zhang and D. Guo, "Wireless peer-to-peer mutual broadcast via sparse recovery," submitted. [Online]. Available: arXiv:1101.0294v1

[6] E. Candès and Y. Plan, "Near-ideal model selection by $\ell_1$ minimization," *Ann. Statist.*, vol. 37, no. 5A, pp. 2145–2177, Oct. 2009.

[7] J. Romberg and R. Neelamani, "Sparse channel separation using random probes," *Inverse Problems*, vol. 26, no. 11, Nov. 2010.

[8] R. Tibshirani, "Regression shrinkage and selection via the lasso," *J. Roy. Statist. Soc. Ser. B*, vol. 58, no. 1, pp. 267–288, 1996.

[9] M. Yuan and Y. Lin, "Model selection and estimation in regression with grouped variables," *J. Roy. Statist. Soc. Ser. B*, pp. 49–67, 2006.

[10] S. Wright, R. Nowak, and M. Figueiredo, "Sparse reconstruction by separable approximation," *IEEE Trans. Signal Processing*, pp. 2479–2493, Jul. 2009.

[11] W. U. Bajwa, "New information processing theory and methods for exploiting sparsity in wireless systems," Ph.D. dissertation, University of Wisconsin, Madison, WI, 2009.

[12] W. Hoeffding, "Probability inequalities for sums of bounded random variables," *J. Amer. Statist. Assoc.*, pp. 13–30, Mar. 1963.

[13] M. Rudelson and R. Vershynin, "Non-asymptotic theory of random matrices: Extreme singular values," in *Proc. Int. Congr. of Mathematicians*, Hyderabad, India, Aug. 2010.

[14] F. MacWilliams and N. Sloane, *The Theory of Error-Correcting Codes*. New York: North-Holland Publishing Company, 1977.

[15] N. Y. Yu and G. Gong, "A new binary sequence family with low correlation and large size," *IEEE Trans. Inform. Theory*, vol. 52, no. 4, pp. 1624–1636, Apr. 2006.

[16] Y. C. Eldar, P. Kuppinger, and H. Bölcskei, "Block-sparse signals: Uncertainty relations and efficient recovery," *IEEE Trans. Inform. Theory*, vol. 58, no. 6, pp. 3042–3054, Jun. 2010.

[17] W. U. Bajwa, R. Calderbank, and S. Jafarpour, "Why Gabor frames? Two fundamental measures of coherence and their role in model selection," *J. Commun. Netw.*, vol. 12, no. 4, pp. 289–307, Aug. 2010.

[18] W. U. Bajwa, R. Calderbank, and M. F. Duarte, "On the conditioning of random block subdictionaries," Department of Computer Science, Duke University, Tech. Rep. TR-2010-06, Sep. 2010. [Online]. Available: http://www.duke.edu/~wb40/pubs/TR2010_block_subdict.pdf

[19] R. A. Horn and C. R. Johnson, *Matrix Analysis*, 1st ed. Cambridge, U.K.: Cambridge University Press, 1985.

[20] J. A. Tropp, "On the conditioning of random subdictionaries," *Appl. Comput. Harmon. Anal.*, vol. 25, no. 1, pp. 1–24, July 2008.

[21] C. Menini and F. V. Oystaeyen, *Abstract Algebra: A Comprehensive Treatment*. New York: Marcel Dekker, 2004.